# Prompting the E-Brushes: Users as Authors in Generative AI

Yiyang Mei[1]

## Abstract

Since its introduction in December 2022, Generative AI (GenAI) has shocked and revolutionized the art world, from winning state art fairs to creating complex videos from simple prompts. Amidst this renaissance, a critical issue arises: should GenAI users be recognized as authors eligible for copyright protection? The Copyright Office, in its March 2023 Guidance, argues against this. Comparing the prompts to clients' instructions for commissioned art, the Office denies users authorship due to their limited control in the creative process. This Article counters this position. It advocates for the recognition of GenAI users who integrate these tools in their creative process. It argues that the current policy overlooks the nuanced and dynamic interaction between GenAI users and the GenAI models, where the users actively shape the output through an iterative process of adjustment, refinement, selection, and arrangement. Instead of excluding the portion of the work that is generated by AI, this Article proposes a simplified and streamlined registration process that acknowledges the use of AI in creation. This approach aligns with the constitutional goal of promoting useful arts and sciences; it encourages people to participate in contributing to the social discourse that is re-fed into the training data; and it shifts the focus from merely pursuing technological advancements to establishing a flexible framework that evolves over time. In conclusion, through a detailed examination of text-to-image generators and the misconceptions surrounding GenAI and user engagement, this Article challenges the Office's view and calls for a regulatory framework that is adaptable o current technological development while ensuring safety and public interest.

---

[1] SJD candidate at Emory University School of Law. Thanks to Matthew Sag for insights and advice and to the participants at the Legal Scholars Roundtable on AI (2024) for their comments and suggestions.

# Table of Contents

## Introduction

Since its release in Dec. 2022, Generative AI (GenAI) has captured wide public fascination. From winning state art fairs,[2] to serving as mental health companions;[3] from passing the theory of mind of a 9-year-old,[4] to generating stunning videos with detailed scenes and complex camera movements based on text prompts,[5] GenAI seems to be starting an AI renaissance in the art world.

But the question is - If a work is created using GenAI, should the user of the model be recognized as the author and granted copyright protection?[6] The Copyright Office has weighed in on this issue. In its guidance issued in March 2023, the Office advises that applications should not include AI-generated content beyond a minimal contribution.[7] The guidance compares the prompts given to AI models to instructions provided by clients to commissioned artists.[8] It suggests that just as commissioned artists use their expertise, preferences, and styles to act on those instructions, AI models leverage their training to produce the work. Since the users of the models don't exert control over the creative process, they shouldn't be granted copyright authorship.

This Article challenges the Copyright Office's position. It argues that the users who significantly and substantively incorporate GenAI tools in their creative process deserve authorship recognition. The current policy, which bars AI-generated content from copyright registration and denies authorship to GenAI model users, stems from a misunderstanding of both the technology and user interaction between models. Contrary to the belief that GenAI operates autonomously or randomly, these are in fact machine learning models that apply complex algorithms to generate outputs closely aligned with input prompts, based on patterns learned from their training data. Users are not mere bystanders in this process – they don't simply input brief descriptions and choose from generated options. Instead, they are actively engaged with the models, using them to discover necessary elements, adjust settings, overcome creative blocks, and even refine their concepts. This integration of GenAI in their workflow helps them overcome obstacles, address inefficiencies, and reduce redundancies. The model users remain the Masterminds of the generated images.

---

[2] *See* Sarah Kuta, *Art Made With Artificial Intelligence Wins at State Fair*, SMITHSONIAN MAGAZINE (September 6, 2022), https://www.smithsonianmag.com/smart-news/artificial-intelligence-art-wins-colorado-state-fair-180980703/

[3] *See* Symposium, Zilin Ma et al., *Understanding the Benefits and Challenges of Using Large Language Model-based Conversational Agents for Mental Well-being Support*, AMIA ANNUAL SYMPOSIUM PROCEEDINGS 1105-1114 (2023)

[4] *See* Bob Yirka, *ChatGPT able to Pass Theory of Mind Test at 9-Year-Old Human Level*, TECH XPLORE (FEBRUARY 17, 2023), https://techxplore.com/news/2023-02-chatgpt-theory-mind-old-human.html

[5] *See* Megan Cerullo, *OpenAI's New Text-to-Video Tool, Sora, Has One Artificial Intelligence Expert "Terrified"*, CBS NEWS (February 16, 2014), https://www.cbsnews.com/news/openai-sora-text-to-video-tool/

[6] *See generally*, Daniel J. Gervais, *The Machine as Author*, 105 IOWA L. REV. 2053 (2020); Shlomit Yanisky-Ravid & Luis Antonio Velez-Hernandez, *Copyrightability of Artworks Produced by Creative Robots and Originality: The Formality-Objective Model*, 19 MINN. J.J. SCI. & TECH. 1, 4-7 (2018); Jane C. Ginsburg & Luke Ali Budiardjo, *Authors and Machines*, 34 BERKELEY TECH. L.J. 343 (2019); Annemarie Bridy, *The Evolution of Authorship: Work Made by Code*, 39 COLUM. J. L. & ARTS 395 (2016); James Grimmelmann, *There's No Such Thing as a Computer-Authored Work - And It's a Good Thing, Too*, 39 COLUM. J. L. & ARTS 403 (2016)

[7] Copyright Registration Guidance: Works Containing Material Generated by Artificial Intelligence, 88 Fed. Reg. 16,190 (Mar. 16, 2023) (to be codified at 37 C.F.R. § 202) ("AI-generated content that is more than de minimis should be explicitly excluded from the application").

[8] *Id.* ("Based on the Office's understanding of the generative AI technologies currently available, users do not exercise ultimate creative control over how such systems interpret prompts and generate material. Instead, these prompts function more like instructions to a commissioned artist – they identify what the prompter wishes to have depicted, but the machine determines how those instructions are implemented in its output.")

As a result, this article suggests that, rather than being required by the Office to identify and exclude AI-generated elements that are more than a minimal amount in the final work, a more streamlined approach should be adopted for registration. The Office could introduce a straightforward checklist in the Application, asking - if the applicants use AI in their creative process, and if they do, if the AI's involvement is substantial and significant? This simplified method would encourage more active engagement with the technology and foster contributions to public discussions, which could then be incorporated back into the training datasets. Such an approach would also align with the constitutional objective of promoting useful arts and sciences. It would set the initial step of establishing a regulatory framework where technology is developing at an ever-increasing pace. It is to recognize, and to emphasize that the essence of law differs fundamentally from technology. The primary role of law is not to chase the latest technological trends but to establish a robust framework that guides the regulation of new advancements for public deployment and safety. This approach demands flexibility rather than a constant, reactive effort to match the pace of technological innovation. In essence, the goal of copyright law is to create a stable foundation that can adapt over time, ensuring that new technologies are integrated into society in a way that prioritizes safety and public interest, without the need for frequent, model-specific adjustments.

Part I provides an introduction to text-to-image generators, explaining their nature and use, and examines the Copyright Office's reaction to this emerging technology. It is divided into three parts - the first part introduces text-to-image generators as a branch of GenAI. Through examples, I will demonstrate how users can modify prompts to refine generated images until they achieve the desired outcome. However, it should be noted that the images created from brief and straightforward prompts lack the depth and detail found in images that have been meticulously edited and adjusted. While this discussion primarily revolves around image generators, it is expected that the insights here will also apply to text-to-video generators, given their shared foundation in Large Language Models (LLMs).

The second part explains the concept of authorship in Copyright law and the Copyright Office's position on the legality of AI-generated content. This analysis will demonstrate that the eligibility for copyright of AI-generated images hinges on their originality – specifically on whether they were independently created by human authors with a minimal degree of creativity. Additionally, I will highlight that the Guidance from the Office outlines seven major themes:

1. There's a two-step process for assessing work created by AI.
2. The Office believes that GenAI is autonomous.
3. The Office believes that GenAI is random.
4. The Office believes that GenAI works as Commissioned Artists.
5. The Office believes that the model users' interactions with GenAI is static and passive; they would only change prompts and select from the generated images one that most aligns with their vision.
6. One can separate the elements created by humans and those generated by the AI.
7. The Office wouldn't register the parts created by AI; the copyright protection would only extend to the parts created by humans.

Based on these findings, I argue that the Copyright Office has four fundamental misunderstandings about GenAI:

1. The Office misunderstands the mechanisms of technology as automatic and random.
2. The Office misunderstands the dynamics between the model users and GenAI models as that between the clients and their commissioned artists.
3. The Office misunderstands the mode of interaction between users and models as passive and static.
4. The Office mistakenly believes that the content created by machines can be clearly distinguished from that created by humans in GenAI-generated images.

Part II and Part III focus on these four misunderstandings. It roughly divides them into two categories - misunderstanding about technology; and misunderstanding about user interaction with technology. Part II specifically addresses the technological aspects, showing that text-to-image generators are not autonomous entities capable of random acts of creation. Instead, their mechanics are highly methodical and structured; their reliance on large datasets and sophisticated algorithms make it very difficult to transcend beyond what is in the dataset. This section aims to dispel the myth that these generators can independently devise plans or create content. It emphasizes the human authors' systematic and iterative efforts in shaping, editing, refining and adjusting the output.

For Part III, the focus is on how users engage with the technology. The purpose is to challenge the notion that users have a passive role in the creative process. Through three examples of a traditional artist using Stable Diffusion in his workflow, a contemporary artist Suanneze Treister's exploration of technology, and Jason Allen's self-narrative using GenAI, this section illustrates that model users have significant control and are actively engaged in shaping the final output. Unlike the clients of commissioned artists who only describe on a high level what they would expect from the artists, model users are in a continuous process of adjustment and refinement in response to the models' generated outcomes.

Part IV proposes that to simplify the registration process for works involving AI-generated materials, the Office should simply ask the applicants whether they have used this tool extensively in their creation process. By acknowledging the interactive nature of the technology, it would also be consistent with the constitutional goal of promoting useful arts and sciences. It would encourage users to use the technology to shape the public discourses that eventually become training data to influence others. Most importantly, it would serve as the first important step to start discussion about regulations of GenAI - to acknowledge that the purpose of law isn't to be most up to date about the latest model in the technology field, but to come up with regulations that can adapt to the changing times.

# I.   Text to Image Generator and the Copyright Office's Response

This section introduces text-to-image generators. It discusses what they are and how people use it. It also explains the Copyright Office's stance on AI-generated images as outlined in their Guidance published in March 2023. This section will have three main parts. The first part introduces the technology; the second part reviews the Office's Guidance; the third part examines the application of this Guidance to current case.

In the first section, I present text-to-image generators as a specific category of GenAI. This section will emphasize the models' abilities to transform textual descriptions to complex images. It then explores the ways in which users engage with these tools, showing that by providing detailed prompts and adjustments, the model users improve and fine-tune the images produced to align with their conceptualizations and visions.

The second part of this section focuses on the Guidance issued by the Copyright Office concerning this phenomenon. I conduct a thematic analysis of the Guidance, arguing that the Office has misunderstood both how text-to-image generators operate, and how model users incorporate them in their workflow. I suggest that, contrary to the Office's assumption that users simply generate complex images through prompts and selection, the model users actually use these tools in a manner much like traditional artists, using them to achieve their visions.

In the final part, I will apply the Guidance to current cases. It will show that the Office emphasizes human authorship and believes that model users don't have control over the creative process when they integrate the tools in their workflow. Merely editing prompts or selecting pictures from generated images don't constitute authorship.

## A. **Text-to-Image Generator**

This part introduces text-to-image generators as a specialized subset of GenAI. It then discusses the capability of these generators to refine images through detailed and progressive adjustments to prompts. The aim here is to demonstrate that the images produced from basic prompts lack the depth of user creativity and control; even minor tweaks to the descriptions can lead to entirely new creations, requiring authors to carefully edit the output to align with their original vision. Furthermore, the images submitted for registrations are highly complex in compositions, stylistic choices, and themes. To create them, the model users must exercise significant control over the creative process.

### a. Text-to-Image Generators are a Subset of Generative AI

Text-to-Image generators are a specialized branch of GenAI, a technology that is capable of producing diverse forms of content, including text, images, audio and synthetic data.[9] Unlike traditional systems that rely on predefined rules and historical data analysis for decision-making and predictions, GenAI is interactive - it generates content based on the prompts users put in. As of October 2023, GenAI includes: text-to-image, text-to-3D, image-to-text, text-to-video, text-to-audio, text-to-text, text-to-code, text-to-science, and others.[10] The focus here, text-to-image, is but one category that transforms written descriptions into visual representations.

### b. How to Use it and What It Produces

---

[9]   *See* George Lawton, *What is Generative AI? Everything You Need to Know*, TechTarget, https://www.techtarget.com/searchenterpriseai/definition/generative-AI

[10] *See* Roberto Gozalo-Brizuela & Eduardo C. Garrido-Merchan, *ChatGPT is Not All you Need. A State of the Art Review of Large Generative AI Models* (Jan. 11, 2022) (unpublished transcript), https://arxiv.org/pdf/2301.04655.pdf?mibextid=Zxz2cZ.

Using text-to-image generators like DALL-E and MidJourney is quite straightforward. DALL-E3, created by OpenAI, is part of the advanced ChatGPT-4 model,[11] which merges LLM with text-to-image capabilities.[12] Users can enter a brief prompt and the model will produce an image. However, the simplicity of these generated images stands in contrast to the work that Jason Allen submits for registration for the Space Theater Opera series. When comparing these images side by side, it becomes clear that even refined prompts fall short of the complexity found in carefully crafted and edited images like Allen's. Therefore, the Copyright Office is incorrect in assuming that images generated by these models are the result of a wholly automated and independent process, and that such detailed works can be produced merely through basic selection and arrangement. It's crucial to recognize that the use of generative models doesn't diminish the creativity and effort that go into producing these images.

**1. Model Users can Generate Images Based on Prompts**

To create images with generative models, model users can put in written descriptions of the image they have in mind. For example, if I want to use ChatGPT-4 to generate an image of an American man using ChatGPT-4, I could write a prompt such as, "draw me an American man."

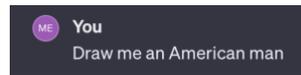

This is what ChatGPT-4 generated:

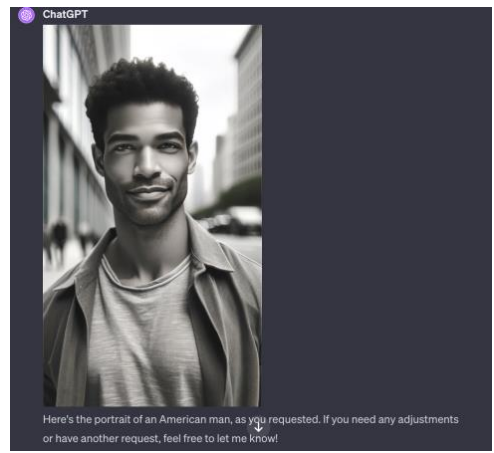

The model determined details like hairstyle, clothing, background, facial expression, and age.

**2. Model Users can Change the Prompts to find the Desired Effects**

But suppose I don't like the picture above and would like to change it. I can do so through editing the prompt:

---

[11] ChatGPT4 includes DALL-E3. *See DALL-E3 is Now Available in ChatGPT Plus and Enterprise*, OPENAI.COM (October 19, 2023), https://openai.com/blog/dall-e-3-is-now-available-in-chatgpt-plus-and-enterprise

[12] *See ChatGPT*, OPENAI.COM https://openai.com/chatgpt

> **ME**  You
> not American enough. Also don't draw in black and white. Give me colored pictures

This is what the model generates:

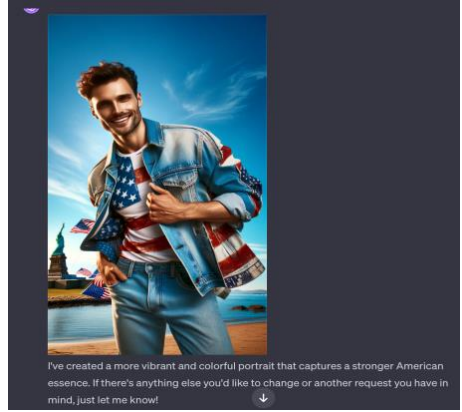

The entire picture changes. But suppose I'd like to make it even more American. I can say:

> **ME**  You
> more American!

ChatGPT-4:

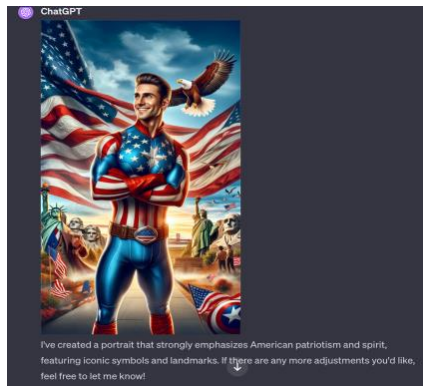

Even more so:

> **ME**  You
> even more American!

ChatGPT-4:

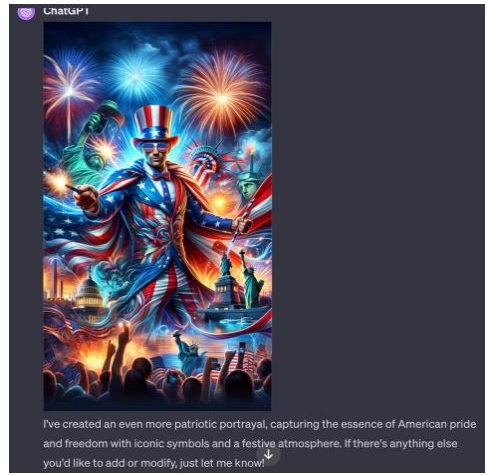

Now, we have an image depicting a man embodying all the quintessential elements of American culture. The only aspect I change during the editing process is the thematic portrayal of "Americanness." But the model changes everything – from composition to the physique, age, and even the race of the man.

The same principle applies to other prompts. Changing them will completely alter the picture. It's nearly impossible to keep the image consistent without personal hands-on editing:

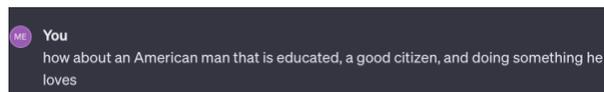

ChatGPT-4:

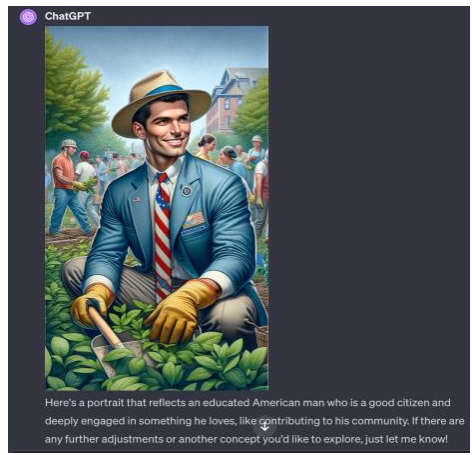

Therefore, it becomes obvious that it is nearly impossible to keep the picture consistent throughout the editing process. The generative models don't just change isolated elements through each iteration. They change the entire picture. To refine the picture through prompts, one would also need to use other tools such as Photoshop and Adobe.

**3. Images Generated by Short Descriptions do not Match the Compositional Complexity of Images that are Carefully Edited**

Editing and refining simple prompts don't create complex images. Compare the images produced above with those submitted for registration. It is obvious that the latter requires much more manipulation of the tool. For example, see Jason Allen's Space Opera Theater:[13]

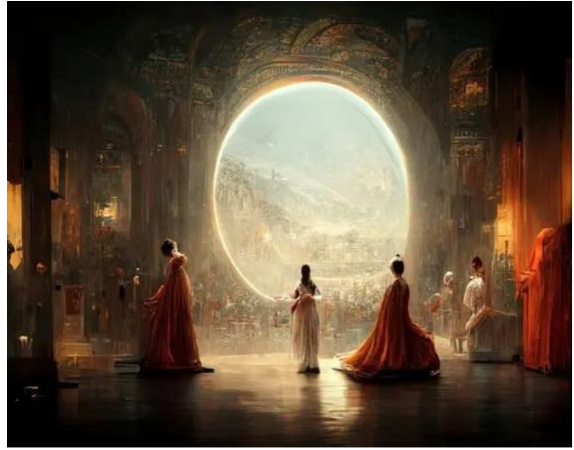

This image is significantly more complex and vibrant than the two examples shown above in terms of concept, composition, intent, and the message conveyed to the viewer. Furthermore, one could also see that the author of the final picture spends considerably more time, energy, and creativity on it compared to the pictures above. As a result, this Article argues that the Copyright Office shouldn't take a blanket rejection of all AI-generated images. It is incorrect to assume that every picture produced by AI is the outcome of a process that is autonomous and automatic.

## B. What is Authorship and the Copyright Office's Response

In March 2023, as a response to the copyright issue posed by GenAI models, the Copyright Office launched an agency-wide initiative to investigate three key questions: 1) whether AI-generated content qualifies for copyright protection, 2) whether works that combine human creativity and AI contributions are eligible for registration, and 3) what details applicants must submit when seeking to register such works?[14] This analysis seeks to analyze the Guidance in depth, highlighting its potential limitations and setting the stage for future discussions about the nature of the technology and its integration into artists' creative process.

This section will be divided into four subsections. The first subsection will explain the concept of authorship under U.S. Copyright law, suggesting that for AI-generated images to be protected by copyright law, the users of the model must independently create the image, contributing more than minimal creativity. The second subsection revisits historical debates on authorship in the contexts of computer-generated works from the 20th century, showing that such discussions have been ongoing since the introduction of the computers.

---

[13] *See* John Wenzel, *A Copyright Battle Over AI-Generated Art will Begin in Colorado*, The Denver Post (March 21, 2023), https://www.denverpost.com/2023/03/21/artificial-intelligence-ai-art-trademark-fight-jason-allen-colorado/

[14] Copyright Registration Guidance: Works Containing Material Generated by Artificial Intelligence, 88 Fed. Reg. 16,190 (Mar. 16, 2023) (to be codified at 37 C.F.R. § 202).

In the third subsection, I will provide an in-depth examination of the Guidance published by the Copyright Office, with the analysis structured in two phases. Initially, I will present a summary of the Guidance, outlining its scope, objectives, concepts and rationales. Following that, I will undertake a thematic analysis of the document. I have identified seven key themes articulated by the Copyright Office: 1) The recommendation of a two-step process for evaluating AI-generated works, 2) the perception of GenAI as autonomous, 3) the characterization of GenAI as random, 4) the analogy of GenAI to commissioned artists, 5) the static nature of user interactions with GenAI, 6) the possibility of distinguishing between human and AI contributions in the final product, 7) the policy against registering machine-produced works.

In the fourth subsection, I argue that the Office has four major misconceptions about this technology – its operational mechanism, its unchanging interaction with users, the flawed comparison between commissioned artists and the models, and its perceived isolation from human-produced works. Each point will be addressed separately in the subsequent sections.

## a. What is Authorship in Copyright

To explore the issue of whether the model user is the author of the generated material, understanding what authorship is is indispensable. This subsection will clarify the concept of authorship under U.S. Copyright Law, focusing on three key components: fixation, works of authorship, and originality. Specifically, it argues that the personality theory of copyright endorses the view that incorporating GenAI in the creative process enhances an author's ability to express themselves. It will also demonstrate that AI-generated images readily fulfill the criteria for works of authorship and fixation. The primary challenge, however, lies in establishing originality, which requires proving that the work was independently created by the human author with a modicum of creativity.

### 1. Personality Theory Supports Using AI for Enhanced Self-Expression by Authors

US copyright law protects original, creative works. Art. I, Section 8, Clause 8 of the Constitution authorizes Congress to provide authors and inventors with exclusive rights to their creations for limited periods.[15] Broadly speaking, copyright is grounded in two main theories: the labor theory and the personality theory.[16] The personality theory, in particular, could be used to advocate for recognizing individuals as the authors of AI-generated works when they actively engage in expressive activities.

The labor theory has been discredited by *Feist Publications, Inc. v. Rural Telephone Service Co.*.[17] Originally, this theory is founded in John Locke's concept of property, which posits that the investment of time, effort, and energy in creating something justifies a claim of ownership over it.[18] As the individuals have spent enough effort in creating the images, they can be recognized as authors of the final generated images due to their contributions to the generative process.[19] *Feist*, however, makes it clear that mere "sweat of the brow"—significant effort—does not qualify for copyright protection.[20] Simply because the individual works towards the final image doesn't mean that they should be the author. Instead, originality is the key factor for copyright eligibility.[21]

---

[15] U.S. CONT. art.1, §8, cl. 8. ("Congress shall have power… to promote the progress of Science and useful Arts, by securing for limited Times to Authors and Inventors the exclusive Right to their respective Writings and Discoveries.")

[16] *See generally* Mala Chatterjee, *Lockean Copyright versus Lockean Property*, 12 J. LEGAL ANALYSIS 136 , 137(2020); Yoo, Christopher S., *Rethinking Copyright and Personhood*, ALL FACULTY SCHOLARSHIP 423 (2019)

[17] 499 U.S. 340, 353 (1991) ("decisions of this Court applying the 1909 Act make clear that the statute didn't permit the 'sweat of the brow' approach.")

In contrast, the personality theory supports the idea that when artists integrate these tools into their creative process, they engage in a form of personal expression that is deeply intertwined with their individuality and artistic vision, thus deserving of copyright protection. This idea is influenced by Hegel.[22] It posits that incorporating technology into the creative process still allows for personal expression, thus meriting recognition of authorship. This theory underscores that an individual's creative work is an expression of their will;[23] it advocates for the protection of such original expressions as reflective of their personality, as these expressions, considered property rights, are fundamental to one's autonomy and identity.[24] AI artists, when they use these tools in their workflow, are still expressing themselves. They ought to be recognized as the author based on this theory.

**2. AI-Generated Works Easily Satisfy Two Key Copyright Criteria: Works of Authorship and Fixation.**

Besides the philosophical basis of copyright, Section 102(a) of Title 17 of the U.S. Code defines the criteria for copyright eligibility. It emphasizes three fundamental requirements: originality, classification as works of authorship, and fixation in a tangible medium.[25] AI-generated images readily fulfill the latter two criteria — being classified as works of authorship and achieving fixation — simply by existing as images. The most difficult problem is determining if the generated Work is original.

The generated image fulfills the two requirements of being works of authorship and fixed because, per the statute, the generated picture falls in the fifth category of a "work of authorship" as pictorial, graphic, and sculptural works.[26] The other condition of fixation requires that the work be embodied "in a tangible medium of expression, now known or later developed, from which it can be perceived, reproduced, or otherwise communicated, either directly or with the aid of a machine or device."[27] The work must also have a degree of permanence or stability, allowing it to be perceived, reproduced, or communicated for a period extending beyond transient duration.[28] This fixation may occur before or simultaneously with the work's publication and must have the author's authorization.[29] The term "tangible medium" includes the physical or digital medium that hosts the copyrighted work, suggesting that copyright protection extends beyond the medium itself. Thus, an image generated by GenAI, qualifies as being "fixed" by existing in a tangible medium. The only requirement left, then, is originality.

---

[18] *See* Justin Hughes, *The Philosophy of Intellectual Property*, 77 Geo.L.J. 287, 297 (1988),

[19] Id.

[20] *Feist*, 499 U.S. at 353

[21] *Id.* ("The sine qua non of copyright is originality")

[22] *See* Margaret Jane Radin, *Property and Personhood*, 34 Stanford L.Rev. 957, 977 (1982)

[23] *See* Yoo, Christopher S., *Rethinking Copyright and Personhood*, All Faculty Scholarship 1039, 1050 (2019) (citing Wenwei Guan, *The Poverty of Intellectual Property Philosophy*, 38 H.K. L.J. 359, 361 (2008); Karla M. O'Regan, *Down- loading Personhood: A Hegelian Theory of Copyright Law*, 7 Canadian J.L. & Tech. 1, 5 (2009); Jeanne L. Schroeder, *Unnatural Rights: Hegel and Intellectual Property*, 60 U. Miami L. Rev. 453, 453 (2006)) (discussing Hegel regarded property as playing an essential function in defining a person's personality).

[24] *Id.* ("Under the Hegelian view, property plays a central role in defining a person as a person. Only by establishing a property interest in external objects can the will achieve a concrete existence.")

[25] 17 U.S.C §102(a).

[26] *Id.* (The eight categories are: 1) literary works; 2) musical works, with or without lyrics; 3) dramatic works, including any accompanying music; 4) pantomimes and choreographic works; 5) pictorial, graphic, and sculptural works; 6) motion pictures and other audiovisual works; 7) sound recordings; and 8) architectural works.)

[27] *Id.*

### 3. The Complex Issue of Originality.

The most challenging aspect for determining the copyrightability of AI-generated images is originality. According to the precedent set by *Feist*, for a work to be considered original, it must be independently created and show a minimum level of creativity.[30] As the later sections of the Article discuss, the Copyright Office believes that AI-generated images don't meet this requirement because the model users lack control over the generation process. I argue that this interpretation is incorrect because it is based on misunderstandings of the technology and the dynamics between users and the generative model. This section lays the groundwork for this argument by exploring the evolution of the concept of originality through landmark cases.

Several cases shaped the concept of originality. The Trade-Mark Cases of 1879 laid the groundwork by linking the Constitution's Intellectual Property Clause to originality.[31] They defined original works as those "founded in the creative powers of the mind" or deemed "the fruits of intellectual labor,"[32] typically manifesting as books, prints, and engravings.[33] The 1884 case of *Burrow-Giles Lithographic Co. v. Sarony*[34] took this discussion further. It questioned Congress' power to grant copyright to photographs.[35] In this dispute, photographer Sarony sued a lithographer for replicating his photo of Oscar Wilde.[36] The issue was whether photos were within the bounds of "writings" of the Copyright clause; and whether it was the artist's creativity or the mechanical operation of the camera that contributed to the work.[37] The court ruled in favor of Sarony. It recognized that when the photographer set the scene, he demonstrated creativity through aspects like posing, lighting, and timing.[38] The case set a precedent that although the subject matter may be replicated, reproducing the photograph with artists' own contribution deserves protection; originality lies in the artist's selection and arrangement of elements.

*Alfred Bell & Co., Ltd. v. Catalda Fine Arts, Inc.* further contributed to this discussion by recognizing that copyright could protect distinguishable variations of works from the public domain.[39] In this case,

---

[28] 17 U.S.C. §101.

[29] *Id.*

[30] 499 U.S. at 345 (1991).

[31] 100 U.S. 82. (1879).

[32] 100 U.S. at 94.

[33] *Id.*

[34] 111 U.S. 53 (1884).

[35] 111 U.S. at 56. ("The constitutional question isn't free from difficulty. The eighth section of the first article of the constitution is the great repository of the powers of Congress, and by the eighth clause of that section congress is authorized to promote the progress of science and useful arts, by securing, for limited times to authors and inventors the exclusive right to their respective writings and discoveries.")

[36] *Id.* at 54

[37] *Id.* at 55 (1884)

[38] *Id.* at 60.

[39] 191 F.2d 99, 102 (2d Cir. 1951) ("Accordingly, we were not ignoring the Constitution when we stated that a copy of something in the public domain will support a copyright if it is a distinguishable variation").

Alfred Bell & Co was a British print producer and dealer that had secured United States copyrights in eight mezzotint engravings of certain paintings in the public domain.[40] It brought an action against a dealer in lithographs, Catalda Fine Arts, Inc., that produced and sold color lithographs of the eight mezzotints.[41] The Court ruled in favor of Alfred Bell. It says that "original in reference to a copyrighted work means that the particular work owes its origin to the author; no large measure of novelty is necessary."[42] "As long as the author contributed something more than a merely trivial variation, something recognizably his own," no matter how poor artistically the author's addition, he would be granted authorship if these were his own.[43] Here, because the engraver contributed to his judgment and conception of the variations, the mezzotints in question weren't simple additions, but creative deviations from the original art; they qualified for copyright protection.

The turning point was reached in *Feist*, which established that originality depended on independent creation and a modicum of creativity.[44] Rural Telephone Service Company, Inc. was a public utility that provided telephone services to several communities in northwest Kansas.[45] It was subject to a state regulation that required all telephone companies operating in Kansas to issue annually an updated telephone directory.[46] Rural published a typical telephone directory, consisting of white pages and yellow pages.[47] The white pages listed in alphabetical order the names of Rural's subscribers, together with their towns and telephone numbers.[48] The yellow pages listed Rural's business subscribers alphabetically by category and feature classified advertisements of various sizes.[49] Rural distributed its directory free of charge to its subscribers, but earned revenue by selling yellow pages advertisements.[50]

Feist Publications, Inc. on the other hand, was a publishing company that specialized in area-wide telephone directories.[51] But unlike a typical directory, which covered only a particular calling area, Feist's area-wide directories covered a much larger geographical range.[52] It approached Rural to obtain white pages listings for its area-wide directory.[53] Rural rejected the request.[54] Feist used Rural's white pages listings without Rural's consent.[55] Rural sued for copyright infringement, arguing that Feist, in compiling its own directory, couldn't use the information contained in Rural's white pages.[56] The issue was whether mere alphabetical arrangement of listings in Rural's directory met the threshold of creativity required for copyright protection.[57]

The Court ruled in favor of Feist. It reasoned that, "original, as the term is used in copyright, means only that the work was independently created by the author (as opposed to copied from other works), and that it possessed at least some minimal degree of creativity."[58] Originality didn't mean novelty; a work may be original even though it closely resembled other works so long as the similarity was fortuitous, not the result of copying.[59] While facts weren't copyrightable, as they didn't owe their origin to an act of authorship, the selection and arrangement of them possessed the requisite originality.[60] Here, because Rural's selection of listings lacked the modicum of creativity to transform mere selection into copyrightable expression, as the information it published is the most basic – name, town, and telephone number about each person who applied to it for telephone service, Feist's use of listings could not constitute infringement.[61]

---

[40] *Id.*
[41] *Id.*
[42] *Id.*
[43] 191 F.2d at103.
[44] 499 U.S. at 346.
[45] 499 U.S. at 342.
[46] *Id.*
[47] *Id.*
[48] *Id.*
[49] *Id.*
[50] *Id.*
[51] *Id.*
[52] 499 U.S. at 342-343.
[53] 499 U.S. at 343.
[54] *Id.*
[55] *Id.*
[56] 499 U.S. at 344.
[57] *Id.*

Applying these rules in the context of AI-generated images, the issue becomes - when model users integrate GenAI models in their workflow, do they independently create the images and show a modicum of creativity? Do the model users exert enough control over the creation process?

### b. Prior Report Advocates that Authorship Belongs to Individuals, not Machines

The question of who the author of the computer-generated content should be wasn't new.[62] Earlier discussions in the 70s concluded that humans should be recognized as the authors of such works.[63] In this article, I argue that this should still be the case.

When Copyright Office received registration applications for works like computer-generated music and abstract art, as well as for projects that incorporated computer-generated elements in 1965, it considered questions such as - should the "work" be considered primarily a result of human creativity, with the computer merely acting as a tool, or did the core aspects of authorship—namely, the idea and its realization—belonged to the computer?[64]

Congress addressed this issue by affirming the principle of human authorship. In 1974, the National Commission on New Technological Uses of Copyrighted Works (CONTU) clarified that the user of the computer was the true author.[65] They argued that it was illogical to ascribe authorship to a computer, likening it to passive devices like cameras or typewriters, which only operated under human control.[66] CONTU maintained that a computer merely followed instructions and lacked the inherent ability to create that was central to authorship.[67] As I further discuss in this Article, this perspective should still prevail despite the recent developments in AI.

---

[58] 499 U.S. at 346.

[59] 499 U.S. at 345.

[60] 499 U.S. at 341.

[61] 499 U.S. at 362.

[62] *See* Edward Lee, *Prompting Progress: Authorship in the Age of AI*, FLORIDA L.REV. (forthcoming 2024); Evan H. Farr, *Copyrightability of Computer-Created Works*, 15 RUTGERS COMPUTER & TECH. L.J. 63, 65 (1989); Pamela Samuelson, *Allocating Ownership Rights in Computer-Generated Works*, 47 U. PITT. L. REV. 1185 (1986); Timothy L. Butler, *Can a Computer be an Author-Copyright Aspects of Artificial Intelligence*, 4 HASTINGS COMM. & ENT. L.J. 707 (1981); Arthur R. Miller, *Copyright Protection for Computer Programs, Databases, and Computer-Generated Works: Is Anything New Since CONTU?*, 106 HARV. L. REV. 977 (1993); Karl F. Jr. Milde, *Can a Computer Be and Author or an Inventor*, 51 J. PATENT OFF. SOC'Y 378 (1969).

[63] *See* CONTU at 45.

[64] U.S. Copyright Office, Sixty-Eighth Annual Report of the Register of Copyrights for the Fiscal Year Ending Jun. 30, 1965, at 5, https://www.copyright.gov/reports/annual/archive/ar-1965.pdf.

[65] *See* CONTU at 45

[66] *See* CONTU at 44.

[67] *Id.*

### c. The Seven Themes of the Current Guidance by Copyright Office

When the Office received applications and inquiries about AI-generated works, it published a Guidance for Works Containing Materials Generated by AI.[68] This section provides a detailed examination of the Guidance. I will do so by first offering an overview of the document, then, I will conduct a thematic analysis of the content. For this guidance, I have identified seven key themes: 1) a two-step process for evaluating works generated by AI, 2) the autonomy of Generative AI, 3) the randomness inherent in GenAI outputs, 4) the role of GenAI as commissioned artists, 5) the static nature of user interactions with GenAI, without iterative processes, 6) the possibility of distinguishing between human-authored and AI-generated components in the final work, and 7) the Copyright Office's policy of not registering works solely produced by machines.

### 1. Overview of the Guidance

In March 2023, the Copyright Office started a comprehensive review after receiving applications for AI-generated content.[69] This initiative aims to explore several key issues: 1) the copyright eligibility of materials created by GenAI, 2) the registrability of works that incorporate both human and AI contributions, and 3) the details applicants must provide when seeking registration for such works.[70] But unlike the CONTU report mentioned above which recognizes the computer as a tool, and the human as the author, the Guidance mentions that if a work's traditional elements of authorship are produced by a machine, the work lacks human authorship and the Office will not register it.[71]

When evaluating submissions that combines human creativity with materials generated by or with the assistance of technology, the Office has established a procedure to determine if a work primarily shows human authorship, using technology as a supportive tool, or if the fundamental aspects of its creation are "conceived and executed by a machine rather than a human."[72] In the case of AI-generated content, the Office would examine whether these elements are merely "results of mechanical reproduction" or they originate from an author's "own original mental conception, to which the author gave visible form."[73]

The Guidance further explains that when the "traditional" elements of authorship in a work are entirely machine-generated, such as when AI creates content in response to a prompt, the work is considered to lack human authorship and is ineligible for registration.[74] This decision is reasonable because the users lack final creative control over the system's interpretation of prompts and the creation of content.[75] The Office compares user prompts to directions given to a commissioned artist – although the prompt may describe the outcome, it is the machine that independently decides the execution of it.[76] The models autonomously decide elements of the picture such as style, pattern, and structure.[77] As a result, the generated image doesn't originate from human creativity; it doesn't qualify for copyright protection.[78]

---

[68] Copyright Registration Guidance: Works Containing Material Generated by Artificial Intelligence, 88 Fed. Reg. 16,190 (Mar. 16, 2023) (to be codified at 37 C.F.R. § 202)

[69] *Id.*

[70] *Id.* ("These technologies, often described as generative AI, raise questions about whether the material they produce is protected by copyright, whether works consisting of both human-authored and AI-generated material may be registered, and what information should be provided to the Office by applicants seeking to register them."

[71] *Id.*

[72] *Id.*

[73] *Id.*

[74] *Id.*

[75] *Id.*

That being said, the Office does recognize a scenario where AI-generated content, if sufficiently combined with human creativity, may qualify for copyright protection. This situation happens when a human artist intentionally selects or organizes AI-created content in a way that makes a composition an original piece of authorship.[79] The second instance is when an artist substantially alters content originally produced by AI, and these alterations meet the copyright criteria.[80] However, it's important to emphasize here the copyright in question here would apply only to the elements created by humans; it doesn't apply to that created by the generative models.[81] And the human-created parts must be "distinct from" and "not influence" the content generated by AI.[82]

## 2. Thematic Analysis of the Guidance

In the Guidance, seven themes are obvious. At the beginning, it introduces a two-step process to assess works that integrate human creativity with AI-generated elements, emphasizing the need to determine the level of human contribution.[83] Then, it discusses the autonomy of AI, questioning whether the technology serves merely as an assisting tool or takes a more active role in the creative process. The Office also explores the idea of AI randomness, proposing these GenAI models are capable of conceiving and executing plans for works of art.[84] The Guidance compares GenAI to commissioned artists, showing that users lack control and creative input when they incorporate GenAI in their work process.[85] It also criticizes the nature of user interactions with AI. Furthermore, it recognizes the potential to distinguish between human-authored and AI-generated components within a work.[86] The document concludes by suggesting that the Office wouldn't register machine-produced work, underscoring the importance of human authorship for copyright eligibility.[87]

---

[76] *Id.*

[77] *Id.* ("These prompts function more like instructions to a commissioned artist—they identify what the prompter wishes to have depicted, but the machine determines how those instructions are implemented in its output. For example, if a user instructs a text-generating technology to "write a poem about copyright law in the style of William Shakespeare," she can expect the system to generate text that is recognizable as a poem, mentions copyright, and resembles Shakespeare's style. But the technology will decide the rhyming pattern, the words in each line, and the structure of the text.)

[78] *Id.* ("When an AI technology determines the expressive elements of its output, the generated material is not the product of human authorship. As a result, that material is not protected by copyright and must be disclaimed in a registration application.")

[79] *Id.* ("In other case, however, a work containing AI-generated material will also contain sufficient human authorship to support a copyright claim. For example, a human may select or arrange AI-generated material in a sufficiently creative way that "the resulting work as a whole constitutes an original work of authorship.")

[80] *Id.* ("Or an artist may modify material originally generated by AI technology to such a degree that the modifications meet the standard for copyright protection.")

[81] *Id.* ("In these cases, copyright will only protect the human-authored aspects of the work, which are "independent of" and do "not affect" the copyright status of the AI-generated material itself.")

[82] *Id.*

[83] *Id.* ("It begins by asking "whether the 'work' is basically one of human authorship, with the computer [or other device] merely being an assisting instrument, or whether the traditional elements of authorship in the work (literary, artistic, or musical expression or elements of selection, arrangement, etc.) were actually conceived and executed not by man but by a machine. In the case of works containing AI-generated material, the Office will consider whether the AI contributions are the result of "mechanical reproduction" or instead of an author's "own original mental conception, to which [the author] gave visible form."")

[84] *Id.*

[85] *Id.* ("Instead, these prompts function more like instructions to a commissioned artist…they identify what the prompter wishes to have depicted, but the machine determines how those instructions are implemented in its output")

**Two-Step Process for Assessing Works**: The Copyright Office outlines a two-step process for evaluating works created using AI. According to the Guidance, the initial inquiry by the Office focuses on whether the work fundamentally reflects human creativity, with technology playing a supportive role, or if the essential elements of authorship—be it literary, artistic, or musical expression, as well as aspects of selection and arrangement—originate from a machine. The document states, it "*begins by asking "whether the 'work' is basically one of human authorship, with the computer [or other device] merely being an assisting instrument, or whether the traditional elements of authorship in the work (literary, artistic, or musical expression or elements of selection, arrangement, etc.) were actually conceived and executed not by man but by a machine.*"[88] For works incorporating AI-generated content, the Office examines "*whether the AI contributions are the result of mechanical reproductions or instead of an author's own original mental conception, to which the author gave visible form.*" The first step would be to establish if the human author's role includes selecting and arranging expressive elements, which makes the work qualifies as human authored. In the second step, if AI plays are used during creation, the question focuses on whether these AI contributions can be attributed to human ingenuity or are merely automated outputs of AI.

**AI autonomy.** The Office believes that text-to-image generators are autonomous. When assessing the eligibility of copyright for AI-created materials, it asks if the computer is "*assisting*," "*conceiving*," and "*executing*" the creative process.[89] To "conceive" and "execute" means that the generative models could independently formulate and execute plans, going from a merely supportive role to actively originating and developing ideas. The Guidance explicitly states, it "*will not register works produced by a machine or mere mechanical possess that operates randomly or automatically without any creative input or intervention from a human author.*"[90] This statement reflects the Office's belief that machines, even in the absence of human creative guidance, have the potential to generate images or content spontaneously.

**AI Randomness.** The Office confirms its belief that these AI models' mechanism could be random when it refuses the to register works that lack intentional human involvement – "*to qualify as a work of authorship, a work must be created by a human being and that it will not register works produced by a machine or mere mechanical process that operates randomly or automatically without any creative input or intervention from a human author.*"[91]

**GenAI as Commissioned Artists**. The Office compares text-to-image generators to commissioned artists in its Guidance, noting that "*based on the Office's understanding of the generative AI technologies currently available, users don't exercise ultimate creative control over how much systems interpret prompts and generate material. Instead, these prompts function more like instructions to a commissioned artist, they identify what the prompter wishes to have depicted, but the machine determines how those instructions are implemented in its output.*"[92] The generative models, according to the Office, don't merely follow instructions from the users; they improve the

---

[86] *Id.* ("copyright will only protect the human-authored aspects of the work, which are independent of and do not affect the copyright status of the AI-generated material itself.")

[87] *Id.* ("If a work's traditional elements of authorship were produced by a machine, the work lacks human authorship and the Office will not register it")

[88] *Id.*

[89] *Id.*

[90] *Id.*

[91] *Id.*

[92] *Id.*

Works by leveraging their pre-programmed preferences and stylistic choices as commissioned artists. For instance, *"if a user instructs a text-generating technology to 'write a poem about copyright law in the style of William Shakespeare,' she can expect the system to generate text that is recognizable as a poem, mentions copyright, and resembles Shakespeare's style. But the technology will decide the rhyming pattern, the words in each line, and the structure of the text."*[93] It is the generative models, rather than the users, that shapes the artistic expression of the work.

**Assumptions about User Interactions with AI**. The Office offers a relatively static and fixed form of interaction between the user and the AI model. It states, *"users do not exercise ultimate creative control over how such systems interpret prompts and generate material…they identify what the prompter wishes to have depicted, but the machine determines how those instructions are implemented in its output.*[94]*" "This implies a scenario where the user provides a concise prompt, and the AI autonomously decides on the visual attributes such as color, composition, and style."*[95] By suggesting that the generative models determine all stylistic choices of the image, the Office assumes that the users won't modify, refine, or alter the generated content subsequently. The users are portrayed as a passive group of people who accept the AI models' creation without further intervention or attempts to change the outcome.

**Separation between AI-generated and Human Created Elements.** The Office believes it is possible to isolate elements in the Work that are created by humans from those generated by AI. It suggests that it will only protect the parts that are created by humans. The Guidance notes, *"copyright will only protect the human-authored aspects of the work, which are 'independent of' and do 'not affect' the copyright status of the AI-generated material itself."*[96]

**Non-registration of Machine-Produced Works**. The Office refuses to grant copyright protection to works produced by machines. It says, *"when an AI technology determines the expressive elements of its output, the generated material is not the product of human authorship. As a result, that material is not protected by copyright and must be disclaimed in a registration application."*[97]

However, the Office concedes that a person may select or change the AI-generated content in a way that satisfies the copyright criteria - *"A human may select or arrange AI-generated material in a sufficiently creative way that the resulting work as a whole constitutes an original work of authorship."*[98] Consequently, the copyright status of a work created using AI is determined on a *"case-by-case inquiry,"*[99] depending on how the generative tool is used and what role it plays in the creative process.

**3. Issues with the Guidance**

The Guidance has four significant misconceptions about the mechanism of the technology, the dynamics between the users and the generative models, the manner in which users engage with these models, and the characteristics of the images produced by AI. Each of these misunderstandings will be addressed here and in more detail in later parts of the Article.

---

[93] *Id.*
[94] *Id.*
[95] *Id.*
[96] *Id.*
[97] *Id.*
[98] *Id.*
[99] *Id.*

**Misunderstanding about the nature of the technology.** The Guidance misunderstands the capabilities of technology by suggesting that it can "conceive" and "execute" plans autonomously. Being able to "conceive" [100] and "execute" [101] means that the machines possess the ability to independently formulate and implement strategies; they have a level of cognitive function like humans. This is incorrect. It's important to recognize that these systems are generative models built on machine learning algorithms and trained datasets. [102] They lack intentional awareness and consciousness as humans. [103] The subsequent section of this Article clarifies this further.

The Office also mistakenly assumes that the generative process is inherently random. To believe so would be attributing the characteristic of one element in the training process – "the noise," to the entire generative process. As I will discuss next, the process of generation is controlled by probabilistic distributions and defined by specific parameters, rather than being purely random or solely dependent on chance. For instance, in systems like Midjourney, consistency can be achieved; using the same prompt will yield the same image, implying a structured and predictable process.

**Misunderstanding about the dynamics between the users and the GenAI models.** The Office's comparison of the dynamics between model users and AI models to the relationship between clients and commissioned artists is inaccurate. Unlike commissioned artists, who leverage their own expertise, preferences, and interpretations into a project, users of generative models maintain control over the end result. Users can fine-tune these models to closely align with their specific vision, while clients would not monitor every step of the creation of the artwork. For instance, in replicating a work like *St. Matthew and the Angel*, a model user would need to precisely dictate the style, ambiance, lighting, texture, and other expressive elements to mirror the original work. In contrast, the commissioned artist Caravaggio would bring his unique naturalistic style and dramatic use of light and shadow to the interpretation. It is the commissioned artists that have control over the image than the clients, unlike for generative art, it is the model users that determine the outcome.

**Misunderstanding about how users use the models.** The Guidance implies an oversimplified mode of interaction between users and generative models. It believes that the model users can put in a basic prompt describing the most rudimentary vision and the AI model will autonomously produce a complex output. During this process, the model will determine the style, format, color, and composition of the picture. This understanding has several issues: 1) it ignores the possibility of the users' writing detailed and comprehensive prompts that show their precise visions and thought process. 2) It overlooks the likelihood of users engaging in iterative processes, refining the output and using other tools such as Photoshop on the outcome until it eventually aligns with their vision. 3) It also presupposes a static nature of interaction where the user simply accepts what the model produces

---

[100] Merriam-Webster, https://www.merriam-webster.com/dictionary/conceive (to form a conception of / to apprehend by reason or imagination.)

[101] Merriam-Webster, https://www.merriam-webster.com/dictionary/execute (to carry something out fully; to put (something) completely into effect; to do what is provided or required by (something))

[102] *What is Generative Model*, DataCamp (Aug 2023), https://www.datacamp.com/blog/what-is-a-generative-model# ("A generative model is a type of machine learning model that aims to learn the underlying patterns or distributions of data in order to generate new, similar data.")

[103] *See generally,* David J. Chalmer, *Could a Large Language Model be Concious?* (March 4, 2023) (unpublished manuscript). https://arxiv.org/abs/2303.07103. (He argues that given mainstream assumptions in the science of consciousness, there are significant obstacles to consciousness in current models: for example, their lack of recurrent processing, a global workspace, and unified agency)

without editing and revision. As the sections below will show, it is this dynamic style of interaction between the model user and the generative models that is the reality.

**Misunderstanding about the characteristics of the images produced by GenAI.** The Office believes that it can separate and isolate the elements created by humans and those generated by machines while evaluating images. It proposes that copyright protection should only be limited to the contributions made by humans.

This suggestion of separating human and machine contributions within a single work of art is impractical. In a work created using AI, it's reasonable to assume that there are elements that the artist will change and others that they decide to keep. The elements that are modified are those that don't align with the artist's vision; those that meet the expectations of the artists and form into a harmonious whole with the rest of the picture would remain untouched. The artists' decision to keep or get rid of the elements are driven more by the overall coherence and harmony of the image, rather than whether certain parts are generated by the AI.[104] Therefore, excluding the elements that are generated by AI in a coherent picture mischaracterizes the artistic creative process.

It's possible that the Office's belief that the generative models can conceive art and operate independently is greatly encouraged by the excitement in social media and news outlets after ChatGPT's release. This excitement fuels speculation that AI might soon equal, or even outpace human capabilities.[105] For example, a few commentaries by the RAND research institute raise alarms about AI's potential existential threats,[106] particularly after a study in May 2023 suggested that LLM have developed a theory of mind surpassing that of a 9-year-old.[107] This idea is further supported by discussions on whether algorithms could outperform human judges in making fair decisions,[108] leading to debates about the possibility of AI replacing human judges.[109] Here, our goal is to get behind the social buzz, establish a standard on the question of copyright authorship against which future technological advancements can be assessed, while ensuring that we understand the nature and capabilities of the technology.

---

[104] *See* J. Derek Lomas & Haian Xue, *Harmony in Design: A Synthesis of Literature from Classical Philosophy, the Sciences, Economics, and Design*, She Ji: The Journal of Design, Economics, and Innovation (2022) https://www.sciencedirect.com/science/article/pii/S240587262200003X (showing that as a philosophy of design, harmony has been used to explain aesthetic beauty, ethical actions, just political systems and sustainable designs; harmony has mathematical properties that emerge in many physical systems. Quoting Christopher Alexander for saying: 'as architects, builders, and artists, we are called upon constantly - every moment of the working day - to make judgments about relative harmony. We are constantly trying to make decisions about what is better and what is worse.'"

[105] *See* David Hamilton & The Associated Press, *The "Godfather of AI" Says He's Scared Tech Will Get Smarter Than Humans: How Do We Survive That?* Fortune (May 4, 2023, 12:06 PM), https://fortune.com/2023/05/04/geoffrey-hinton-godfather-ai-tech-will-get-smarter-than-humans-chatgpt/ (Hinton mentioned in an interview with MIT Technology Review that he has suddenly switched his views on whether these things are going to be more intelligent than humans; he says that he thinks they are very close to it now and they will be much more intelligent than us in the future); Sue Halpern, *A New Generation of Robots Seems Increasingly Human*, The New Yorker (July 26, 2023) ( https://www.newyorker.com/tech/annals-of-technology/a-new-generation-of-robots-seems-increasingly-human. *See contra.* Mika Koivisto & Simone Grassini, *Best Humans Still Outperform AI in a Creative Divergent Thinking Task*, Scientific Reports 13 (2023), https://www.nature.com/articles/s41598-023-40858-3

[106] *See Current Artificial Intelligence Does Not Meaningfully Increase Risk of a Biological Weapons Attack*, RAND (January 25, 2024),https://www.rand.org/news/press/2024/01/25.html (suggesting that just because today's LLMs aren't able to close the knowledge gap needed to facilitate biological weapons attack planning doesn't preclude the possibility that they may be able to in the future); *See also* Edward Geist & Andrew J. Lohn, *How Might AI Affect the Risk of Nuclear War?* RAND (Apr 24, 2018) https://www.rand.org/pubs/perspectives/PE296.html

[107] *See* Michal Kosinski, *Theory of Mind May Have Spontaneously Emerged in Large Language Models* (Feb 4, 2023) (unpublished manuscript), https://arxiv.org/abs/2302.02083; *see also* Thilo Hagendorff et al., *Human-Like Intuitive Behavior and Reasoning Biases Emerged in Large Language Models but Disappeared in ChatGPT*, Nature Computational Science 3, 833-838 (2023). https://www.nature.com/articles/s43588-023-00527-x.

[108] *See* Ben Green & Yiling Chen, *Disparate Interactions: An Algorithm-in-the-Loop Analysis of Fairness in Risk Assessments*, FAT*' 19: Proceedings of the Conference on Fairness, Accountability, and Transparency (January 2019), 90-99 https://dl.acm.org/doi/10.1145/3287560.3287563; *See also* H Mahmud et al., *What Influences Algorithmic Decision-Making?*

## d. Case Application

The Copyright Office applies the Guidance to three current cases: *Thaler v. Perlmutter*,[110] a copyright ruling concerning the *Zarya of the Dawn*,[111] and a second request for reconsideration concerning the refusal to register *Theater d'Opera Spatial*.[112] As I show below, in all three cases, the Office emphasizes the importance of human authorship, underscoring that direct generation from single-line prompt isn't sufficient for copyright eligibility. The Office further explains that given the unpredictable nature of text-to-image generators, the model users don't exercise sufficient control over the tool to warrant authorship.

### 1. Thaler v. Perlmutter

Stephen Thaler is a computer scientist. He develops and owns AI programs which he claims as being capable of generating original pieces of visual art, much like the output of a human artist.[113] One of his systems here is the "Creativity Machine," which he uses to create the work "A Recent Entrance to Paradise."[114]

Thaler attempted to register this work with the Copyright Office.[115] In his application, he put the Creativity Machine as the Author, explaining that the image was "autonomously created by a computer algorithm running on a machine."[116] However, he argued that he should still have the copyright of the

---

*A Systematic Literature Review On Algorithm Aversion*, Technological Forecasting and Social Change (February 2022) https://www.sciencedirect.com/science/article/pii/S0040162521008210 (Finding that algorithms consistently outperform humans in decision-making)

[109] *See* Richard M. Re & Alicia Solow-Niederman, *Developing Artificially Intelligent Justice*, 22 Stan. Tech. L. Rev. 242 (2019); Tim Wu, *Will Artificial Intelligence Eat the Law? The Rise of Hybrid Social-Ordering Systems*, 119 Colum. L. Rev. 2001 (2019); Zichun Xu, *Human Judges in the Era of Artificial Intelligence: Challenges and Opportunities*, Applied Artificial Intelligence (2022) https://www.tandfonline.com/doi/full/10.1080/08839514.2021.2013652 (arguing that judging from the current judicial application practice, it seems inevitable that judges will be replaced by artificial intelligence.); Jimmy Hoover, *Chief Justice Roberts: AI Won't Replace Human Judges*, Law.com (December 31, 2023 at 6:00 PM), https://www.law.com/nationallawjournal/2023/12/31/chief-justice-roberts-ai-wont-replace-human-judges/#:~:text=contacts%20you%20provided.-,Chief%20Justice%20John%20Roberts%20Jr.,heart%20of%20the%20court%20system. (saying that Justice Roberts wrote in his 2023 annual year-end report that AI will have a profound impact on the jobs of judges but won't replace the fundamentally human discretion at the heart of the court system).

[110] *See* No. 22-1564 (BAH), 2023 WL 5333236, *4-*6 (D.D.C. 2023).

[111] U.S. Copyright Office Letter to Lindberg *re: Zarya of the Dawn* (Registration # VAu001480196) (Feb. 21, 2023), https://www.copyright.gov/docs/zarya-of-the-dawn.pdf [hereinafter *Zarya of the Dawn* decision]

[112] Reply from the U.S. Copyright Office about Second Request for Reconsideration for Refusal to Register Theatre d'Opera Spatial (SR # 1-11743923581; Correspondence ID: 1-5T5320R), To Tamara Pester, Esq. (September 5, 2023) https://www.copyright.gov/rulings-filings/review-board/docs/Theatre-Dopera-Spatial.pdf [hereinafter *Theatre d'Opera Spatial* decision]

[113] No. CV 22-1564 (BAH), at *1 (D.D.C. Aug. 18, 2023)

[114] *Id.*

[115] *Id.*

[116] *Id.*

"computer-generated work" himself as a work-for-hire to the owner of the Creativity Machine.[117] The Copyright Office denied the application, concluding that this particular work will not support a claim to copyright, because the work lacked human authorship.[118]

The Court confirmed the Copyright Office's decision. It held that while we were approaching new frontiers in copyright as artists put AI in their toolbox to be used in the generation of new visual and other artistic works,[119] the fact that "authorship" was synonymous with human creation had persisted since copyright law had otherwise evolved.[120] This case wasn't nearly so complex — as the machine generated the image based on one line of instruction, the machine determined the composition and stylistic choices. Thaler didn't exercise authorial control over the creation of the picture.[121] He shouldn't have human authorship.[122]

## 2. Copyright Ruling re Zarya of the Dawn

Kristina Kashtanova is an AI educator and consultant.[123] They used MidJourney to create a comic book. The book consists of 18 pages, one of which is a cover.[124] The cover page consists of an image of a young woman, the title of the Book, and the words "Kashtanova" and "Midjourney" (see below).[125] The remaining 17 pages consist of mixed text and visual material.[126] In the application, they list the author of the work as "Kristina Kashtanova."[127]

The Copyright Office recognized Kashtanova as the author of the text and the compilation, which included the work's written and visual elements. This was because Kashtanova wrote the text by themselves, without the assistance of any other source of tool, including any GenAI program.[128] For the compilation - they selected, refined, cropped, positioned, framed, and arranged the images to create the story told within its images.[129]

---

[117] *Id.*
[118] *Id.*
[119] *Id.* at 6.
[120] *Id.* at 5.
[121] *Id.* at 6.
[122] ID.
[123] Kristina Kashtanova identifies as non-binary, pronoun they/them. They have used GenAI to create a series of AI art such as arya of the Dawn, Rose Enigma, Burning Man, Videos - Animations. *See* Kris Kashtanova, INSTAGRAM (March 8, 2022), https://www.instagram.com/p/Ca2xcDZlo2z/?img_index=2; Kris Kashtanova, *Portfolio*, https://www.kris.art/
[124] U.S. Copyright Office Letter to Lindberg re: *Zarya of the Dawn* (Registration # VAu001480196) (Feb. 21, 2023), https://www.copyright.gov/docs/zarya-of-the-dawn.pdf [hereinafter *Zarya of the Dawn* decision].
[125] *Id.*
[126] *Id.*
[127] *Id.*
[128] *Id.* at 4 ("The Office agrees that the text of the Work is protected by copyright. Your letter states that "the text of the Work was written entirely by Kashtanova without the help of any other source or tool, including any generative AI program." Based on this statement, the Office finds that the text is the product of human authorship.")
[129] *Id.* at 5 ("The Office also agrees that the selection and arrangement of the images and text in the Work are protectable as a compilation…Ms. Kashtanova states that she "selected, refined, cropped, positioned, framed, and arranged" the images in the Work to create the story told within its pages…Based on the representation that the selection and arrangement of the images in the Work was done entirely by Ms. Kashtanova, the Office concludes that it is the product of human authorship.")

As to the images that were generated by Midjourney, Kashtanova didn't have copyright over them.[130] Based on the Copyright Office's understanding, Midjourney generated four images based on one line of prompt.[131] As Midjourney didn't interpret prompts as specific instructions to create a particular expressive result, and it couldn't understand grammar, sentence structure, or words like humans,[132] the process by which a Midjourny user obtained an ultimate satisfactory image through the tool wasn't the same as that of a human artist, writer, or photographer.[133] By only "guiding" the structure and content of each image, Kashatanova didn't create the traditional elements of authorship in the images.[134] They didn't control Midjourney as a tool to reach their desired images; Midjourney generated it in an unpredictable way.[135] Kashatanova lacked sufficient control over generated images to be treated as the author behind them.[136]

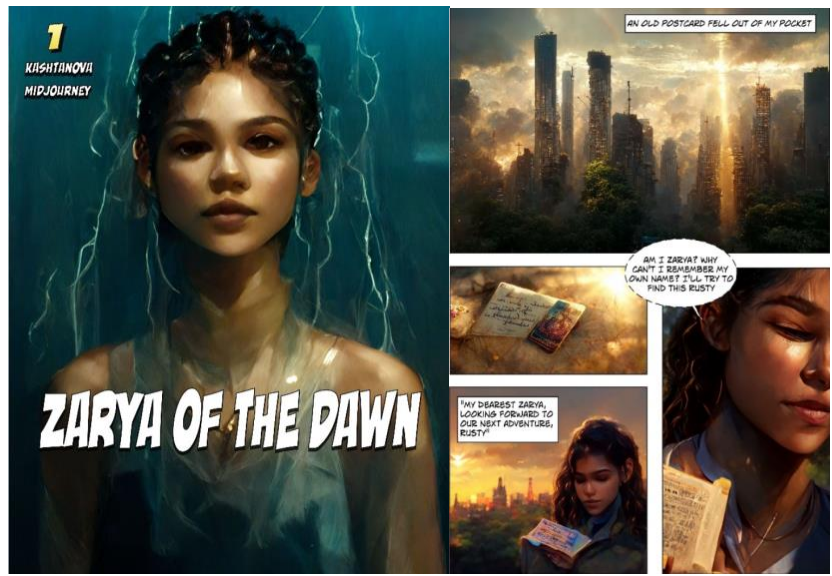

### 3. Second Request for Reconsideration for Refusal to Register Théâtre D'opéra Spatial

The third example is Jason Allen's picture. After he created the Space Opera Theater series, Allen tried to register the Work. Because the Office knew that AI-generated material contributed to the Work, the examiner assigned to the application requested additional information about Allen's use of Midjourney.[137] In response, Allen provided an explanation of his process, stating that he put in numerous revisions and text prompts at least 624 times to arrive at the initial version of the image.[138] He further explained that after Midjourney produced the initial version of the Work, he used Adobe Photoshop to remove flaws and create new visual content and used Giapixel AI to "upscale" the image, increasing its resolution and size.[139]

---

[130] *Id.* at 8.
[131] *Id.* at 7.
[132] *Id.*
[133] *Id.* at 8.
[134] *Id.* (The Office says that though she claims to have "guided" the structure and content of each image, the process described in Kashtanova Letter makes clear that it was Midjourney - not Kashtanova - that originated the "traditional elements of authorship" in the images.)
[135] *Id.* at 9.
[136] *Id.* at 9-10 (The Office suggests that because Midjourney starts with randomly generated noise that evolves into a final image, there's no guarantee that a particular prompt will generate any particular visual output. Kashtanova's using Midjourney is like commissioning a visual artist to do the same.)
[137] *Theatre d'Opera Spatial* decision at 2.

The Office rejected his application.[140] Requesting that the feature of the Work generated by Midjourney be excluded from the copyright claim, the Office confirmed that the Work couldn't be registered without limiting the claim to only the copyrightable authorship Allen himself contributed to the Work.[141]

Allen appealed. He argued that his creative input into Midjourney is on par with that expressed by other types of artists.[142] The Office refuted the claim. It noted that Allen's sole contribution to the Midjourney Image was inputting the text prompt that produced it.[143] However, putting in the prompt wasn't enough, as the creation process was still dependent on how Midjourney processed the prompts. Given the technology's nature, it was anticipated that users would go through hundreds of iterations before arriving at an image they found truly satisfactory.[144] Therefore, Allen shouldn't be granted authorship for the work here.

## II. The Copyright Office Misunderstood how the text-to-image generators work

The previous sections argue that the Copyright Office has misconstrued the nature of text-to-image generators, labeling them as "automatic" and attributing to them the ability to conceive plans. This section aims to clarify these misconceptions. It will demonstrate that: 1) generative models are the culmination of years of scientific research and experimentation; they are not entities that have appeared out of nowhere; 2) these models are fundamentally statistical, refining their outputs to match the input text and training data closely; 3) their operations are not random; 4) they do not possess cognitive abilities; and 5) they cannot function independently. I will particularly focus on Stability AI's Stable Diffusion to discuss diffusion models, latent space, and the CLIP model.[145] Additionally, I will introduce OpenAI's latest text-to-video model, Sora, showing that it is the combination of diffusion model, latent space, and the transformer architecture.

---

[138] *Id.*

[139] *Id.*

[140] *Id.*

[141] *Id.*

[142] *Id.* at 3.

[143] *Id.* at 6. ("In the Board's view, Mr. Allen's actions as described do not make him the author of the Midjourney Image because his sole contribution to the Midjourney Image was inputting the text prompt that produced it.")

[144] *Id.* at 7 (Noting that it is the Office's understanding that because Midjourney doesn't treat text prompts as direct instructions, users may need to attempt hundreds of iterations before landing on an image they find satisfactory. This appears to be the case for Mr. Allen)

[145] Although there are three major text-to-image generators like Midjourney, Stable Diffusion, and ChatGPT3 and 4, all share substantial similarities - Midjourney uses a diffusion model. *See* Dev, *How Does MidJourney Create Images in Real Time?* MEDIUM (Oct 20, 2023) https://medium.com/hackrlife/how-does-midjourney-create-images-in-real-time-a07fad2df3da; Stable Diffusion is a latent text-to-image diffusion model that uses a fixed, pretrained text-encoder CLIP. *see Stable Diffusion v2-1-unclip Model Card*, HUGGING FACE, https://huggingface.co/stabilityai/stable-diffusion-2-1-unclip (model description). DALL-E2 uses a stable diffusion model that integrates data from the CLIP model. *See An Introduction to Using DALL-E3: Tip, Examples, and Features*, DATACAMP (Nov 2023) https://www.datacamp.com/tutorial/an-introduction-to-dalle3.

I begin this section by discussing the structures of text-to-image generators. By explaining the core components of this subtype of GenAI, including its pre-processing components and transformers, I show that it behaves in a highly structured and methodical manner, with each component designed to perform specific tasks that contribute to the overall functionality. I then introduce the learning mechanism of the text-to-image generators. I demonstrate that the supervised learning used by DALL–E2, Imagine, and Draw, are constrained by their dataset; the generative models learn to create images that match specific textual descriptions. The training data of the models significantly influences the generative process. This analysis reveals that the generative process is not as unpredictable as the Office suggests.

After discussing the structural and operational aspects of the models, I will conduct a comparative analysis between human creativity and machine processes. This comparison illustrates that, unlike human creators who possess the ability to evaluate and reflect on their creations, GenAI functions solely on computational calculations. They are not capable of evaluating the aesthetics of the outputs as people do.

To further drive home the distinction between generative models and human cognition, and the idea that these models aren't autonomous, but rather, imitations of human outputs that fall short of true understanding and consciousness, I will borrow examples from computational functionalism and John Searle's Chinese Room. These examples seek to refute the idea that simply increasing a model's neuronal counts could replicate human cognitive functions; machine learning models work like the human mind.[146] They will show that 1) the Office's interpretation of the GenAI as autonomous takes a functional view of the human mind, 2) GenAI, despite its advanced capabilities in generating complex outputs, fundamentally lacks the creative intent and adaptability that is intrinsic to human cognition, and 3) mere manipulation of symbols, even when it results in seemingly intelligent language or text, does not imply comprehension or understanding by the program generating such content. The purpose is to reemphasize the point that significant gaps do exist between machine creation and human creativity. The subjective, emotional experience of people cannot be replaced by machines.
This section will be structured into four subparts. The first subpart will examine the structure and learning mechanism of GenAI; the second subpart will review and explain how diffusion models, latent space, and text-image integration work. It argues that none of the current models possesses the capability to conceive ideas and formulate plans. The third subpart focuses on the idea that GenAI, despite their impressive abilities to generate complex images, don't have understanding or comprehension of the output; the manipulation of the symbols doesn't mean that they have intention to create; the Copyright Office's belief that the GenAI models have autonomy may stem from a functional understanding of mental states. The fourth subpart tries to explain why the Copyright Office is reluctant to spend serious efforts to understand and investigate the operation of the generative models. It argues that the Office is influenced by the inherent anthropocentric bias against computer-generated art – that in spite of its recognition of its objective aesthetic value, it doesn't feel connected to the picture. This emotional disconnection, coupled with the assumption that AI art is easily created, contributes to the Copyright Office's reluctance.

---

[146] *See* Anne Trafon, *Study Urges Caution when Comparing Neutral Networks to the Brain*, MIT News (November 2, 2022), https://news.mit.edu/2022/neural-networks-brain-function-1102 (Saying that in the field of neuroscience, researchers often use neural networks to try to model the same kind of tasks that the brain performs, in hopes that the models could suggest new hypotheses regarding how the brain itself performs those tasks. However, a group of researchers at MIT is urging that more caution should be taken when interpreting these models.)

## A. The Generative process of Text-to-Image Generators is Methodical.

The mechanism of text-to-image generators is systematic and methodical. It involves a structured sequence of steps, starting with pre-processing, followed by generation, and ending with ongoing refinement and adjustments. This systematic approach is based on supervised learning, meaning that what the model generates is based on the data it's trained on. Unlike humans, who can critically assess their creations, machines lack the ability to make such evaluative judgments. Instead, they operate within the boundaries of pre-set models and functions, adhering to a predictable process.

### a. Refining, Adjusting, and Evaluating: The Iterative Process of Text-to-Image Generators

Text-to-image generators use a systematic approach that incorporates data pre-processing and advanced models like transformers.[147] These transformers are essential for interpreting text descriptions accurately.[148] The process is carefully organized, with each component having a distinct role that contributes to the system's overall effectiveness. For example, in models such as DALL-E2 and DALL-E3, transformers are not only used for understanding the text but also for encoding it in a way that directs the image generation process.[149] This ensures that the generated images closely match the textual descriptions. More specifically, transformer models convert text into a format that text-to-image applications can understand, focusing on important parts of the text to create coherent images through self-attention mechanisms.[150] This structured method of understanding and generation demonstrates the predictability of the image generation process, challenging the misconception that the resulting images are produced randomly.

[147] *See* Luvv Aggarwal, *Data Preprocessing for the GenAI Project*, MEDIUM (Jul 16, 2023), https://medium.com/@luvvaggarwal2002/data-preprocessing-for-the-genai-project-2b982d4275f0 (Saying that data preprocessing is a crucial step in the machine learning pipeline. Raw data is often messy; it often has missing values, inconsistencies, and noise that can negatively impact the performance of predictive models. So one needs to manipulate, filter, or augment the data before it's analyzed. It includes essential steps like encoding and normalization. Proper encoding of categorical variables and text data allows the model to process and understand the information effectively; normalization ensures that numeric features are scaled appropriately, preventing features from dominating others during model training); *see also* Itzikr, *Generative AI- Aguide on Data Preparation*, ITZIKR'S BLOG (July 7, 2023), https://volumes.blog/2023/07/07/generative-ai-a-guide-on-data-preparation/; Marko Vidrih, *New AI Breakthrough - Google's Muse: Text-to-Image Generation via Masked Generative Transformers*, MEDIUM (Jan 16, 2023), https://vidrihmarko.medium.com/new-ai-breakthrough-googles-muse-text-to-image-generation-via-masked-generative-transformers-6bedd5b0cad9 (introducing that Muse: Text-to-Image Generation via Masked Generative Transformers is a promising new technology); *see also* Ryan O'Connor, *How DALL·E2 Actually Works*, ASSEMBLYAI (SEP 29, 2023), https://www.assemblyai.com/blog/how-dall-e-2-actually-works/ (Suggesting that DALL-E2 uses transformer model)
[148] *See* MissGorgeousTech, *Transformers: How AI is Learning to Understand Human Language*, MEDIUM (Feb 17, 2023), https://medium.com/the-abcs-of-ai/transformers-how-ai-is-learning-to-understand-human-language-a57995022e90#:~:text=Transformers%20use%20self%2Dattention%20mechanisms,summarization%2C%20and%20question%2Danswering. (suggesting that transformers use self-attention mechanisms to process input text and generate output text. The self-attention mechanism allows the model to focus on different parts of a sentence to better understand what is being said.)
[149] *See DALL-E3*, OPENAI https://openai.com/dall-e-3 (last visited Mar 3, 2024) (introducing that DALL-E3 is built on ChatGPT, which lets the user uses ChatGPT as a brainstorming partner and refiner of the prompts to generate images). *See generally, DALL·E: Creating Images from Text*, OPENAI (last visited Mar 3, 2024) https://openai.com/research/dall-e (ChatGPT, on the other hand, is a pre-trained model built on the transformer architecture.) *See* Amit Prakash, *What is Transformer Architecture and How does it Power ChatGPT?*, THOUGHTSPOT (Feb 23, 2023), https://www.thoughtspot.com/data-trends/ai/what-is-transformer-architecture-chatgpt.
[150] *See* Witold Wydmanski, *What's the Difference Between Self-Attention and Attention in Transformer Architecture?*, MEDIUM (DEC 3, 2022), https://medium.com/mlearning-ai/whats-the-difference-between-self-attention-and-attention-in-transformer-architecture-3780404382f3 (introducing self-attention as the ability of a transformer model to attend to different parts of the same input sequence when making predictions. This mechanism allows us to look at the whole context of the sequence while encoding each of the input elements.)

**b. The Learning Mechanism of Text-to-Image Generators Further Shows that the Generative Process isn't Random. It's Based on What is in the Training Data.**

Text-to-image generators such as DALL-E2 use supervised learning.[151] This model is trained on extensive datasets consisting of labeled text-image pairs, learning to create images that match specific textual descriptions.[152] What is generated isn't random. It's based on what is in the training data.

The way supervised learning works is by initially putting together a dataset made up of input-output pairs, with every input matched to a specific output.[153] The algorithm then uses this training data to establish a relationship between inputs and outputs, fine-tuning its internal parameters to minimize the gap between its predicted outcomes and the actual output labels.[154] Following the training phase, the model undergoes evaluation using a new, previously unseen dataset to test its accuracy and effectiveness.

Another way to think about supervised learning is to compare it to the way a student artist learns to paint. Initially, the student studies some samples; they are guided by detailed instructions that explain the purpose and technique behind each brushstroke and color choice. By repeatedly practicing these master techniques, the student gradually learns to produce new artwork in a similar style. Similarly, in supervised learning, algorithms are trained using datasets filled with examples of existing artwork, each labeled with the correct style. Through training, the algorithms learn to replicate the styles and characteristics of the art in the dataset, much like the student artist who learns to emulate the techniques of the masters they study. Therefore, this learning mechanism isn't random. It follows a kind of predetermined path and pattern that's inherent in the training dataset.[155]

**c. Generative Models vs. Human Creativity: The Missing Link in AI's Evaluative Process**

Generative models differ significantly from human creativity, not only because they follow the underlying patterns of their training data, but also, they are formulaic in their creation process. Human creativity usually involves four stages: preparation, incubation, illumination, and verification.[156] Machines fall short in each of these areas. In contrast to the flexible and ingenious methods humans use, algorithms gather data for preparation, lacking the ability to truly comprehend the content, not engaging with the incubation process of unconscious thinking. Without the capacity for reflection, the algorithm models also miss out on the illumination phase of connecting ideas. Finally, they don't undergo verification, as they lack the crucial evaluative phase that involves reflection and thought.

---

[151] DALL-E2's architecture uses CLIP. *See* Anshu Kumar, *Understanding OpenAI CLIP & Its Applications*, MEDIUM (Nov 19, 2022), https://akgeni.medium.com/understanding-openai-clip-its-applications-452bd214e226 (introducing CLIP using natural language supervision).

[152] *See CLIP: Neural Network Capable of Classifying Images without Prior Training on the Classes*, GOOGLE CLOUD (last visited March 3, 2024)https://console.cloud.google.com/vertex-ai/publishers/google/model-garden/22?pli=1 (suggesting that CLIP is trained on a variety of image-text pairs with the capability of classifying the images into one of several classes).

[153] *See What is Supervised Learning?*, GOOGLE CLOUD https://cloud.google.com/discover/what-is-supervised-learning (last visited Mar 3, 2024) (The data in supervised learning is labeled - meaning that it contains examples of both inputs and correct output. The algorithms analyze a large dataset of these training pairs to infer what a desired output value would be when asked to make a prediction on new data).

[154] *Id.*

[155] This section isn't to claim that all text-to-image generators use supervised learning. It recognizes that there are other models such as Google's Imagen that uses a mix of supervised and unsupervised learning. *See Imagen: Unprecedented Photorealism x Deep Level of Language Understanding*, GOOGLE RESEARCH https://imagen.research.google/ (last visited March 3, 2024) (however, here, it only discusses supervised learning, because the purpose isn't to review the learning mechanisms of all the text-to-image generators. It is only to serve as an example that the three learning mechanisms of supervised, unsupervised, and semi-supervised have a structured path that's not completely random.)

At each stage, human creativity is more nuanced than that of machines. In the preparation stage, for example, individuals define the problem and gather relevant information.[157] This is in sharp contrast to GenAI's approach of simply scraping online data to build a machine learning dataset. During the incubation stage, humans might take breaks or divert their attention, allowing their subconscious to process the problem, which can lead to insightful reflections later.[158] In contrast, LLM, despite their impressive ability to generate text, lacks the capability for such subconscious processing or genuine comprehension of the content they produce, let alone reflections or evaluations.

The third stage, illumination, characterized by sudden "Eureka" moments, presents a significant challenge for machines. For humans, these moments can lead to revolutionary insights that go beyond existing knowledge.[159] Take Antoine Lavoisier as an example. He questioned the then-prevailing phlogiston theory, which claimed that a fire-like element called phlogiston was released during combustion, making objects lighter. Lavoisier designed experiments to demonstrate that in sealed containers, the mass of a substance remains the same before and after combustion. This discovery established a new chemical paradigm, highlighting the vital role of oxygen in combustion.[160] However, such innovative concepts, not present in any dataset, are inaccessible to GenAI. It can't go beyond its dataset. Relying on GenAI for scientific development wouldn't work as it cannot create ideas outside its existing knowledge base.

The final phase, verification, poses yet another challenge for machines. This phase involves testing, refining, and perfecting ideas – processes that are naturally intuitive and reflective, and beyond the capabilities of a generative model. For humans, this stage requires a critical evaluation of their work, encouraging them to reconsider their approaches.[161] Machines, on the other hand, prioritize technical precision and the application of algorithms, concentrating solely on reducing errors. Unlike humans, they lack the ability to contemplate and assess their decisions.

---

[156] I used Wallas' as an example because his is the earliest and most foundational upon which other theories are built on. *See* Maria Popova, *The Art of Thought: A Pioneering 1926 Model of the Four Stages of Creativity*, THE MARGINALIAN (last visited March 3, 2024); *see also* Marion Botella et al., *What Are the Stages of the Creative Process? What Visual Art Students Are Saying*, FRONTIERS IN PSYCHOLOGY (2018) https://www.ncbi.nlm.nih.gov/pmc/articles/PMC6259352/ (listing 20 other theories of creativity).

[157] *See* Marion Botella et al., *What Are the Stages of the Creative Process? What Visual Art Students Are Saying*, FRONTIERS IN PSYCHOLOGY (2018), https://www.ncbi.nlm.nih.gov/pmc/articles/PMC6259352/ (in the first stage of preparation, individuals define the problem and gather information in order to solve it).

[158] *Id.* (Incubation is a time of solitude and relaxation, where idea associations take place at a subconscious level.)

[159] *Id.* (The individual experiences an illumination or insight with the emergence of an idea, an image or a solution.)

[160] In the mid-18th century, the most pressing issue in chemistry was to determine what exactly happens when something burns. The prevailing theory was that flammable materials contained a substance called "phlogiston" that was released during combustion. The theory held that when a candle burned, for example, phlogiston was transferred from it to the surrounding air. When the air became saturated with phlogiston and could contain no more, the flame went out. *The Chemical Revolution of Antoine-Laurent Lavoisier*, ACS CHEMISTRY FOR LIFE https://www.acs.org/education/whatischemistry/landmarks/lavoisier.html#:~:text=The%20prevailing%20theory%20was%20that,it%20to%20the%20surrounding%20air (last visited March 3, 2024).

[161] Marion Botella et al., *What Are the Stages of the Creative Process? What Visual Art Students Are Saying*, FRONTIERS IN PSYCHOLOGY (2018) https://www.ncbi.nlm.nih.gov/pmc/articles/PMC6259352/ (for the verification stage, new ideas are tested and verified, leading to the elaboration of a solution and to its production).

## B. Understanding the Mechanics of GenAI - An In-Depth Exploration of Text-to-Image and Text-to-Video Generators

Unlike what the Copyright Office suggests in the Guidance, text-to-image generators do not have the ability to conceive plans or originate thoughts about artwork. They are machine learning models that generate outcomes consistent with textual input, relying entirely on patterns found in their training data.[162]

The purpose of this section is to provide a clear understanding of the three fundamental concepts in image generation: latent space, diffusion model, and CLIP. It will establish that these foundational concepts and structures are grounded in a rigorous mathematical framework. Additionally, this section will illustrate how the latest advancement, Sora, integrates these foundational concepts for text-to-video generations. Discussing these models show that in spite of their dramatic entrance into the public domain in 2022, they are the culmination of extensive development and training. They aren't enveloped in mysteries or possess the potential to wreak havoc to human civilization through superhuman intelligence.

This section will be divided into five parts. The first part provides a high-level overview of how Stable Diffusion works. It focuses on explaining the three concepts of latent space, diffusion model, and CLIP. The second part delves into the concept of latent space, particularly its application in Stable Diffusion. It will explain latent space as a method that simplifies complex entities into manageable elements for enhanced processing and analysis. It suggests that the origin of this method can be traced to Pearson, the father of statistics.

The third part focuses on the diffusion model. It introduces the model as a method that incrementally deconstructs images into a chaotic distribution and then methodically reconstructs them to match a given input. The process isn't random – it involves a series of steps that gradually add and remove noise, effectively "learning" how to create complex patterns from simpler forms.

In the fourth part, I will explore the integration of text and image, underlining how this approach highlights the mathematical foundation of these models. The implementation of CLIP has significantly improved the alignment between text inputs and visual outputs. This discussion serves as a reminder that, regardless of their complexity, these models do not have human-like creativity or intelligence.

The fifth and final part introduces Sora, the latest advancement in text-to-video modeling. It posits that despite Sora's remarkable capabilities in generating and simulating videos, it essentially is built on principles of diffusion, transformers, and latent diffusion, concepts previously discussed. It further reiterates the point that these technologies are marked by evolutionary and gradual progression rather than a revolutionary leap.

---

[162] *See Learn About Generative AI,* Google Search Help https://support.google.com/websearch/answer/13954172?hl=en#:~:text=Generative%20AI%20is%0a%20type,just%20great%20at%20finding%20patterns. (last visited March 3, 2024) (Introducing GenAI as a type of machine learning model; saying that it's not a human being. It can't think for itself or feel emotions. It' just great at finding patterns.)

**a. Three Concepts in Text-to-Image Generation: Latent Space, Diffusion Models, and CLIP**

The discussion in this section focuses on the types, architecture, and mechanisms of text-to-image generators. It shows that, for Stable Diffusion, for example, there are primarily three fundamental concepts at work: latent space, diffusion model, and CLIP. All three are built on mathematical models and decades of previous research. In simple terms, as discussed below, latent space is a multi-dimensional space where each dimension represents some feature learned from the data. It's like an artist's palette, where every hue and shade imaginable is at one's fingertips, waiting to be combined into a masterpiece. However, within this space, the images are not yet formed; they are mere possibilities awaiting creation. For the diffusion model, it generates images by starting with a pattern of random noise and gradually shaping it into coherent images over multiple steps. They are like artists who start with a canvas full of random splashes and colors. Over time, they try to shape these colors and strokes into a recognizable picture with their own style.

CLIP marks a watershed moment in text to image generation. It is a method that bridges the gap between visual concept and natural language. It empowers the models to generate images that more closely align with the nuances of language, even taking in account of contexts in sentence. It works like a critic who understands both pictures and words. Before CLIP, the artist could misinterpret what the painting would convey to the audience. But with CLIP, the artist is equipped with an official explanation of the meaning of the picture. It allows that the image aligns much more closely with the story they are supposed to teach.

The way the three concepts work in Stable Diffusion is this – when the user enters a prompt in the generator, the model leverages CLIP to find a point in the latent space that corresponds to the text description. Then, the diffusion model works to progressively reduce the noise and refine the abstract representation back to a coherent image that aligns with the prompt. This reverse diffusion process is guided by the information encoded in the latent space provided by CLIP (Figure 1).[163]

Understanding these three concepts and the image generation process will show that the creation process of the generators is rooted deeply in the mechanical analysis of data distributions. It is also worth noting here that the purpose of discussing technology isn't to be most up to date about technology – a task more suited to a systematic review – but to extend beyond the discussion of technology, and to underscore the difference between humans and GenAI. It is to highlight that these generative models cannot be compared to humans.

---

[163] Jay Alammar, *The Illustrated Stable Diffusion*, GitHub (Nov 2022), http://jalammar.github.io/illustrated-stable-diffusion/ (Illustrating that Stable Diffusion is a system made up of several components. It's not one monolithic model. There's a text-understanding component that translates the text information into a numeric representation that captures the ideas in the text, and then the information is presented to the Image Generator. The text encoder is a special Transformer language model CLIP which takes the input text and outputs a list of numbers representing each word/token. Then, the diffusion model (UNet + Scheduler) produces an information array that the image decoder uses to paint the final image).

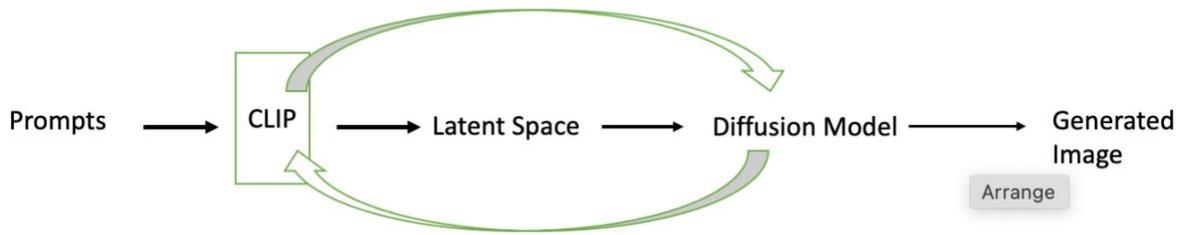

Figure 1

## b. Latent Space in Stable Diffusion

Latent space is a complex yet systematic concept rooted in historical statistical methods - particularly in Pearson's work, rather than being arbitrary as suggested by the Copyright Office. It simplifies images into basic forms for efficient and detailed image generation. This section focuses on how latent space plays its part in Stable Diffusion and its development.

At its core, latent space is crucial to the operation of Stable Diffusion, which merges the two ideas of the diffusion model and latent space.[164] This synergy enhances image generation while optimizing computational resources. The concept was introduced by Rombach et al. in their seminal paper *High-Resolution Image Synthesis with Latent Diffusion Models*.[165] In that paper, they use latent space within autoencoders to decompose complex images into basic forms, then use the diffusion model to generate or alter images.[166] Imagine starting with a complex image you wish to recreate or use as a base for new creations. The proposed method simplifies this image to its basic outlines, stripping away all but the essential details. Then, much like an artist filling in a sketch, the image is gradually enriched with details, step by step, until it reemerges as a rich, full picture. This two-step process of simplification and elaboration allows computers to generate images more efficiently and with greater detail.

The initial simplification occurs in latent space, which serves as a method to reduce complex scenarios to a manageable form and iteratively refine this approximation. This approach, which seems new to the public, actually revisits an old concept traceable to Karl Pearson. Pearson is the "father of statistics."[167] He introduced Principal Component Analysis (PCA) as a dimensionality reduction technique to preserve maximum information in a multivariate dataset.[168] PCA essentially sorts through data with numerous variables to identify and simplify patterns, allowing for easier understanding and use of the data.[169] It helps focus on the most crucial elements within a complex picture, allowing us to concentrate on the significant details without distraction.

---

[164] *Id.*

[165] Robin Rombach et al., *High-Resolution Image Synthesis with Latent Diffusion Models* (Apr 13, 2022) (unpublished manuscript), https://arxiv.org/pdf/2112.10752.pdf.

[166] *Id.*(In Methods section, they suggested that to circumvent the drawback of high computational demands of diffusion models towards high-resolution image synthesis, they used an autoencoding model which learns a space that is perceptually equivalent to the image space, but offers significantly reduced computational complexity).

[167] *See* Bernard J. Norton, *Karl Pearson and Statistics: The Social Origins of Scientific Innovation*, Social Studies of Science, Vol. 8, No. 1 (1978), https://faculty.fiu.edu/~blissl/Pearson1.pdf.

[168] Generally, PCA is a statistical method used for reducing the dimensionality of complex datasets. It's particularly helpful in projects with many variables, where not all variables are equally important. It allows the analyst to identify and focus on the primary key variables, reducing the less critical ones. It involves several steps: 1) standardization - adjusting the variables to a standard scale, 2) covariance matrix computation – identifying the relationships between different variables, 3) feature

PCA is also a form of unsupervised learning that relies entirely on the input data itself without reference to the corresponding target data.[170] Understanding unsupervised learning is important because it is a key machine learning technique in GenAI.[171] It allows algorithms to learn patterns in unlabelled data, much like a student learning to paint by observing various styles without formal instruction, yet gradually grasping these elements independently through exploration.[172]

The key concept in latent space - simplifying complex images into simple chunks proposed in PCA, gained further traction with the introduction of Variational Autoencoders (VAEs) in 2013. VAEs are neural networks designed to make predictions based on unobservable factors.[173] It's like looking for causes behind a few phenomenon in detective fiction. Introduced by Diederik P. Kingma and Max Welling in their work *Auto-Encoding Variation Bayes*, VAEs address the challenge of making predictions in complex data by simplifying situations into more manageable forms and iteratively refining these approximations.[174] This process is like solving a complex puzzle, which starts with grouping puzzle pieces and gradually refining their placement. Without VAEs, the simplification can lead to the "mean field" limitation, where oversimplified data may lose critical details, leading to less accurate or generalized predictions.[175] To address this difficulty, Kingma and Welling introduced a more effective estimator that refines the model's approximations.[176] This new method is built on the wake-sleep algorithm and Stochastic Variational Inference (SVI).[177]

Both wake-sleep algorithms and SVI are mathematical models that rely on probabilistic distributions to aid understanding and predicting underlying patterns by simplifying complex data and optimizing computational load. The wake-sleep algorithm improves probabilistic models by alternating between

---

vectors – determining the principal components, 4) recasting data: aligning the data along the principal components axes to view it in the reduced dimension space. *See A Step-by-Step Complete Guide to Principal Component Analysis | PCA for Beginners* https://www.turing.com/kb/guide-to-principal-component-analysis (last visited March 3, 2024).

[169] *Id.*

[170] PRINCIPAL COMPONENT ANALYSIS (PCA), http://www.stats.org.uk/pca/ (last visited March 3, 2024)

[171] Muneeb Umerani, *The Wonders of GenAI*, MEDIUM (Aug 12, 2023), https://medium.com/@muneebishere2020/the-wonders-of-genai-feaccf79a88#:~:text=The%20reason%20of%20using%20a,a%20subset%20of%20Deep%20Learning.&text=3%2D%20Uses%20supervised%2C%20unsupervised%20and%20semi%2Dsupervised%20methods. (Machine Learning is a subset of AI; ML are of two types, unsupervised with unlabeled data and supervised with labeled data).

[172] *Id.*

[173] J.Rafid Siddiqui, *Latnt Spaces (Part-2): A Simple Guide to Variational Autoencoders*, MEDIUM (Dec 14, 2021), https://medium.com/mlearning-ai/latent-spaces-part-2-a-simple-guide-to-variational-autoencoders-9369b9abd6f (The main strength of autoencoder lies in their ability to extract the abstract representation of the data space which is supposed to handle unseen instances. This allows the model to generate new images that have not already been seen using the latent space. The general autoencoder architecture doesn't allow much freedom in traversing the latent space. This can be circumvented by VAEs which learn a latent distribution instead of a latent vector.)

[174] *See* Diederik P. Kingma & Max Welling, *Auto-Encoding Variational Bayes* (Dec 10, 2022) (unpublished manuscript), https://arxiv.org/abs/1312.6114

[175] *Id.* (Discussing in the introduction that the previous variational Bayesian approach involves the optimization of an approximation to the intractable posterior. Unfortunately, the common mean-field approach requires analytical solutions of expectations, which are intractable in the general case. This paper shows how a reparameterization of the variational lower bound yields a simple differentiable unbiased estimator of the lower bound, i.e., more accurate predictions).

[176] *Id.* at 6.

[177] *Id.*

the "wake" and "sleep" phases.[178] In the "wake" phase, the algorithm learns from real-world data. For instance, it might observe an image of a crowded street and identify different elements like cars and pedestrians. During this phase, it also makes guesses about latent variables - the unseen factors influencing what it observes.[179] Then, in the "sleep" phase, the model uses its learning to create new scenarios, refining its initial guesses.[180] This cycle, much like unsupervised learning, enhances the model's ability to understand and predict underlying patterns in data.

SVI, or Stochastic Variational Inference, serves to streamline complex data analysis. Technically, it simplifies complex posterior distributions in probabilistic models into simpler alternatives, thereby enhancing Bayesian computation for handling large datasets.[181] Probabilistic, rather than deterministic models are used here because we'd like to model the uncertainties in the data. Using probabilities allows for a more efficient estimation of these distributions, particularly when the data is huge.

In summary, the concept of latent space is not as novel a concept as the Copyright Office implies in its Guidance. Its theoretical origins go back to the 20th century, and it is intimately related to other concepts of PCA and VAEs. Furthermore, the technique isn't automatic. It is simply a representation of compressed data in which similar data points are closer together in space.[182]

### c. The Methodical Nature of Diffusion Models

A second indispensable concept in Stable Diffusion in the diffusion model. It aids the task of learning the complex relationship between text and images by gradually transforming random noise into detailed visuals that align with given text prompts. As this section will show, the operation of diffusion models is highly structured, not just due to the process itself, but also because of the three key steps involved in data preparation: normalization, range shifting, and ensuring numerical stability.

The diffusion model works in a structured, two-phase process that carefully balances the introduction and removal of noise in data. Initially, it introduces a controlled amount of randomness, which is like layering a clear image with digital "dust" that progressively obscures the details.[183] This process isn't a random degradation but a carefully calibrated one. The subsequent phase involves reversing this process, where the model removes the digital "dust" to restore the image's clarity.[184] This cycle of adding and then removing noise is crucial, as it compels the model to focus on the data's key features, thereby preventing overfitting and enhancing its ability to generalize to new data.

---

[178] *See* Geoffrey E Hinton et al., *The Wake-Sleep Algorithm for Unsupervised Neutral Networks*, Science vol 268, Issue 5214 (1995), https://www.cs.toronto.edu/~hinton/absps/ws.pdf. (Introducing this algorithm as an unsupervised learning algorithm for a multilayer network of stochastic neuron. In the "wake" phase, neurons are driven by recognition connections, and generative connections are adapted to increase the probability that they would reconstruct the correct activity vector in the layer below. In the "sleep" phase, neurons are driven by generative connections. The recognition connections are adapted to increase the probability that they would produce the correct activity vector in the layer above)

[179] *Id.*

[180] *Id.*

[181] *See* Andy Jones, *Stochastic Variational Inference*, GitHub, https://andrewcharlesjones.github.io/journal/svi.html (last visited March 2024) (introducing Variational Inference as a framework to approximate intractable posterior distributions. Stochastic Variational Inference is a family of methods that uses stochastic optimization techniques to speed up variational approaches and scale them to large datasets). *See also* Hoffman et al., *Stochastic Variational Inference* (Apr 22, 2013) (unpublished manuscript), https://arxiv.org/pdf/1206.7051.pdf.

[182] *See* Ekin Tiu, *Understanding Latent Space in Machine Learning*, Medium (Feb 4, 2020), https://towardsdatascience.com/understanding-latent-space-in-machine-learning-de5a7c687d8d.

[183] *See* Kemal Erdem, *Step by Step Visual Introduction to Diffusion Models*, Medium (Nov 9, 2023),https://medium.com/@kemalpiro/step-by-step-visual-introduction-to-diffusion-models-235942d2f15c (The diffusion process is split into forward and reverse diffusion processes. The forward diffusion process is a process of turning an image into noise, and the reverse diffusion process is supposed to turn that noise into the image again).

To effectively denoise an image, the programmer must adjust the model's parameters, specifically the mean and variance.[185] This is like managing the flow of water through a network of pipes obstructed by debris. The "noise" acts as the debris, disrupting the information flow. To restore smooth flow, like adjusting water pressure and valve settings (representing the mean and variance), these parameters are fine-tuned to ensure information flows unimpeded through the model.

Supposedly, this approach is borrowed from the concepts in Langevin dynamics – a mathematical framework that describes the motion of particles in a fluid, which account for both deterministic movements and random, Brownian motions.[186] The integration of random forces into the model is pivotal for handling the inherent uncertainty and variability in data, which is a critical aspect of the denoising process. This methodical approach underlines the diffusion model's ability to navigate and mitigate noise through a structured and systematic process, rather than relying on randomness.[187]

Besides training the model, the data preparation stage for the diffusion model is also formulaic. The images, after being scraped from online, would first undergo a notable scaling transformation of normalization to make sure that every feature in the input data, like the pixel values in an image, is treated equally.[188] Typically, they are shifted from the conventional 0 (complete black) -255 (complete white) range to a range between -1 (representing what would've been black) and 1 (representing what would've been white).[189] The purpose is to maintain numerical stability and ensure that the data works well with the processes in a neural network. Then, after the image data is transformed to the new scale, and before the decoder interprets and converts it back again to a form people can understand, the image takes a detour at the Gaussian-distributed latent space, where all values follow the bell-curve pattern centering around 0 and has a spread of 1.[190] Once the values achieve a mathematical harmony, the decoder works to interpret the distribution-based representation.[191]

These tasks: normalization and scaling to make sure all elements are in proportion to each other, shifting range to adjust all elements to a standard measurement system, and maintaining numerical stability to make sure the neural networks are consistent and compatible is highly methodical and structural. It is more about the data being systematically reshaped, adjusted, and optimized based on pre-programmed mathematical rules, rather than the system being random.

In conclusion, the workings of the diffusion model and the data preparation stages in advanced generative models like Stable Diffusion are far from arbitrary or random. These processes are underpinned by rigorous mathematical principles and structured methodologies that guide the transformation and interpretation of data.

---

[184] *Id.*

[185] Naman Rastogi, *Navigating DDPMs - A Closer Look at Denoising Diffusion Probabilistic Models*, Medium (Aug 17, 2023), https://medium.com/@deep_space/navigating-ddpms-a-closer-look-at-denoising-diffusion-probabilistic-models-a55f74d5227a.

[186] *See* Lilian Weng, *What are Diffusion Models*, GitHub, https://lilianweng.github.io/posts/2021-07-11-diffusion-models/ (last visited March 3, 2024); *See also* Jonathan Ho. et al., *Denoising Diffusion Probabilistic Models* (Dec 16, 2020) (unpublished manuscript) at 8 https://arxiv.org/pdf/2006.11239.pdf.

[187] *Id.* (Section 3. Diffusion models and denoising autoencoders.)

[188] *Id.* at 4.

[189] *Id.*

[190] *Id.* at 4-5

[191] *Id.*

### d. Bridging Text and Image: CLIP

The sections above explain how Stable Diffusion uses concepts like latent space and diffusion models to compress, represent, and generate images. As a review, latent space refers to a multi-dimensional environment where data tokens are condensed to identify underlying patterns. This is followed by a denoising process to enhance image clarity. Contrary to what some may believe, this procedure is not arbitrary. It is grounded in mathematical principles and relies on probability distributions. The diffusion model introduces noise to blur images, then systematically removes this noise to clarify them. This methodical approach of alternating between adding and removing noise is carefully designed to prevent overfitting, allowing the model to apply knowledge from known datasets to new, unseen ones. The seemingly random "noise" can be explained by Brownian motion, a concept from physics, underscoring that the process is far from haphazard or coincidental, despite what the official guidance might suggest.

In this section, I discuss CLIP. As mentioned above, Stable Diffusion uses a CLIP trained encoder to convert text to embeddings, meaning that the natural language is converted to numerical representations ("embeddings") that captures the essence of the words.[192] The model is developed by OpenAI, which instead of solely relying on mapping the essential elements of complex images onto a multi-dimensional space, comes up with a method working by integrating text with images to enhance the quality of the generated visuals. This technique draws on a rich tradition in cultural production, finding parallels in comic books, pop art, and conceptual art.[193]

This section is organized into two main parts: The first part explores the approach to text-image integration prior to the introduction of CLIP. It highlights the use of Transformer architecture and Discrete Variational Autoencoders (dVAE) for image compression. It shows that this method involved merging compressed image tokens with text in a stepwise, highly structured process. Additionally, this section contrasts technological image generation with human creativity, noting that while human creativity often thrives on deviating from learned patterns, technological methods remain more constrained and formulaic. The second part focuses on CLIP. It offers a very basic understanding of how it categorizes and understands the pairing between words and images. It shows that the image generation process is strategic and driven by algorithms. It follows a logical sequence, which is much different from the subjective and spontaneous act of human creation. The purpose of this section is to show that: 1) A precursor model exists before the introduction of CLIP, 2) CLIP enhances the capabilities of this earlier model, 3) Despite advancements, both models fundamentally depend on probabilistic distributions and are heavily influenced by the training data they're fed. This underscores that the image generation process is systematic and not left to chance, 4) These models' ability to produce remarkable images hinges on human input and interaction. People are indispensable in the creative process.

---

[192] *See* Aayush Agrawal, *Stable Diffusion Using Hugging Face*, MEDIUM (Nov 9, 2022), https://towardsdatascience.com/stable-diffusion-using-hugging-face-501d8dbdd8.

[193] For example, comic books, a form of media that emerged in the 19th century, have long used a mix of text and images to capture the imagination. Series like Superman, Batman and Robin, Wonder Woman, Plastic Man, and Green Lantern offered an affordable and accessible gateway to fantastical worlds in a world post economic depression. The pop art movement mirrors this trend. Artists like Roy Lichtenstein, Andy Warhol, and Barbara Kruger incorporated text into their art to question social standards and spark reflection among audiences.

**1. Text-Image Integration Before CLIP.**

Before the introduction of CLIP, Ramesh and his team used a different approach that focused on different parts of a sentence to capture the essence of the sentence. They used text to guide the creation of images through a technique known as the Transformer architecture.[194]

The model's training process was mechanical and heavily relied on the architecture and computational abilities of the design. At the first step, the team used a model called a discrete variational autoencoder (dVAE) to simplify large images into more manageable versions.[195] This compression converted a high-resolution 256x256 image into a pixel-art style 32x32 version, where each pixel could assume one of 8192 unique values.[196] Then, at the second step, text was broken down into pieces.[197] These textual tokens, along with their image counterparts, formed a stream that was then fed into a transformer with 12 billion parameters.[198] These parameters are fine-tuned to optimize the model's performance. Then, after processing this stream of image and text tokens, the transformer learned how the texts and images were related.[199] Think of it this way: imagine you had a huge, detailed painting, and you'd like to explain it to someone. Generally speaking, as a first step, you turned the painting into a smaller, simpler sketch that still captured the essence of the original. Then, once you thought about how you'd describe the sketch to someone, you used words to highlight the important parts and how they fit together. The outline of the painting was paired with the verbal description so that people could understand the picture without seeing the original painting.

This operation is substantially different from how people create paintings - while the human artists create masterpieces precisely because they break from patterns and norms, these generative models excel when they follow the trained patterns. For the artists, the process is not about conforming to learned patterns from simplified pictures but about the capacity to diverge from the previously learned things. Artistic innovation often emerges from challenging the conventional, venturing into uncharted territories, and crafting something truly original. Take, for instance, the vibrant composition of Van Gogh's "Sunflowers," which owes its charm not to adherence to the muted tones of his era but to his bold rejection of them. Similarly, Picasso's work in Cubism stemmed from his deliberate move away from realistic representation, embracing a more abstract, fragmented approach.

This approach of breaking away from the previous model doesn't work for generative models. If it is trained on a dataset consisting of realistic images and subdued colors, it would not generate styles that show vibrant colors and scenes of intense emotional depth with dramatic use of light and shadow. Instead, it would stick to the original pattern. This method of creation that replicates and optimizes within a framework is inherently different from the human creativity that thrives on the very act of breaking free from established norms. As generative models don't have the ability to innovate and to step beyond the familiar bounds, they couldn't be compared to humans' abilities.

---

[194] *See* Aditya Ramesh et al., *Zero-Shot Text-to-Image Generation* (Feb 26, 2021) (unpublished manuscript) https://arxiv.org/pdf/2102.12092.pdf (Describing an approach to generate images based on a transformer that autoregressive models the text and image tokens as a single stream of data. This paper also introduced DALL-E)
[195] *Id.* (Methods section)
[196] *Id.*
[197] *Id.*
[198] *Id.*
[199] *Id.*

## 2. CLIP: A more straightforward Approach

CLIP introduces a more effective approach to align texts with images. Unlike the older method, which required simplifying images before comparison, CLIP directly pairs images with their textual descriptions by enhancing similarities in correct matches and reducing them in incorrect ones.[200] Imagine a vast array of photos and captions needing pairing. Previously, before the introduction of CLIP, one would need to simplify all the photos to make them easier to handle, and then compare the simplified version with the caption to find the best match. CLIP, however, turns this into a competitive game, scoring points for each successful match and deducting for errors. Over time, CLIP gets really good at this game, making it easier and quicker to identify and pair photos with captions.[201] This ongoing process of trial and error, far from being random, is precisely modeled to improve the system's matching capabilities over time, even for the things it hasn't been trained on.

CLIP's effectiveness lies in its instance discrimination techniques as part of contrastive learning abilities. This approach allows the model to differentiate between various pieces of information by comparison.[202] It groups similar items together while distancing the dissimilar ones.[203] Once similar items are clustered, the model takes a closer step by focusing on individual data instances.[204] It examines two modified versions of the same image, which might involve changes like zooming in, reducing sharpness, resizing, or altering colors. Through this comparative analysis, the model learns to recognize the images as identical despite their modifications. This method enhances the model's ability to accurately identify and "understand" the specifications of an image.

In conclusion, CLIP represents a systematic and structural processing of text-image pairs. It underscores that its mechanism isn't automatic and random as suggested by the Copyright Office. Essentially, it is an algorithm working exactly as it is programmed.

## e. SORA

Sora is the latest text-to-videos model. It can generate videos up to a minute long, featuring highly detailed scenes, complex camera motion, and multiple characters with vibrant emotions.[205] It can also create videos based on a still image or extend existing footage with new material. Its architecture relies on four components: a diffusion model, a transformer, patches, and latent space,[206] all of which have been discussed previously. As a result, the underlying message for understanding the mechanism of the technology remains the same: despite its ability to generate impressive images, these are still models fundamentally based on mathematical formulas.

---

[200] *See CLIP: Connecting Text and Images,* OpenAI, https://openai.com/research/clip. (Last visited March 3, 2024).

[201] *See generally,* Radford et al., *Learning Transferable Visual Models From Natural Language Supervision,* https://arxiv.org/pdf/2103.00020.pdf

[202] *See* Shashank Vats, *Unleashing the Potential of Zero-Shot Classification Using OpenAI CLIP* (Feb 26, 2021) (unpublished manuscript), https://medium.com/aimonks/unleashing-the-potential-of-zero-shot-classification-with-contrastive-learning-1d2567ea1b13. (The fundamental idea of contrastive learning is Instance Discrimination - the unlabelled data points are juxtaposed against each other to teach models which points are similar and which are different. Those belonging to the same distribution are pushed towards each other in the embedding space whereas those belonging to different distribution are pulled away)

[203] *Id.*

[204] *Id.*

[205] *See* M Saravanan, *Sora OpenAI: the AI Model that Generates Mind-Blowing Videos From Text,* Medium (Feb 16, 2024) https://medium.com/@iamsaro1996/sora-openai-the-ai-model-that-generates-mind-blowing-videos-from-text-8f2ceda8d900.

## C. The Generative Models Don't Have Human-Like Creativity

The discussions above emphasized the mechanical nature of image and video generators; this section pushes this argument further by clarifying that they don't think and create as humans. The models do not understand the cultural significance of their output, nor can they reflect on the social implication of the generated content. What they have is mere manipulation of the symbols. It would be a mistake to attribute the ability to conceive and to execute to these models.

Specifically, this section is organized in two parts. The first part introduces computational functionalism. It explores the source of the Copyright Office's mistake and argues that it may be swayed by the functionalist approach in cognitive science — because these models can generate content similar to humans, the Office believes they think and work as humans. This is incorrect. As shown by the second part, the models' ability to imitate the outward behavior of people don't imply that they genuinely understand the content. There's a fundamental difference between GenAI and human intelligence; just because artists integrate them in their creative process doesn't mean they lose control over the generated output.

### a. The Office's Functional Understanding of the Models

This section shows that GenAI models aren't capable of genuinely understanding the content. The Office's perspective, which treats these models as independent entities, adopts a functionalist view. It implies that their external behaviors are indicative of their internal states. However, just because these models can produce images via a black-box process doesn't necessarily mean that they have attained autonomy and independence.

The fact that GenAI doesn't have a genuine understanding of their output should be fairly apparent by now. Emily Bender et al. has argued extensively in their seminal paper *On the Dangers of Stochastic Parrots: Can Language Models be Too Big*, that as the models grow in size and complexity, they essentially become "stochastic parrots" adept at mimicking human language patterns without actual comprehension or understanding.[207] The idea is that while they can generate beautiful sentences, the string of words are based on tokens of predictions. They don't come from authentic understanding of contexts. For example, when people try to use LLM for mental companions and to alleviate a sense of loneliness, the output feels rather empty. While the models can generate messages that sound superficially empathetic and inclusive, they are not personalized and thus lacking connections with the audience.

---

[206] *Video Generation Models as World Simulators,* OpenAI, https://openai.com/research/video-generation-models-as-world-simulators (last visited March 3, 2024) (Sora is the combination of diffusion transformer and latent diffusion model. It works by starting with each frame of the video consisting of static noise, and then uses machine learning to gradually transform the images into something resembling the description in the prompt. The part for latent space allows it to turn videos into patches by first compressing videos into a lower-dimensional latent space and subsequently decomposing the representation into spacetime patches. The transformer component allows it to use the attention mechanism to focus on different parts of the sentence. The foundation remains rooted in probabilistic distributions that make the output most aligns with the text and what is in the training data.)

[207] *See* Emily M. Bender et al., *On the Dangers of Stochastic Parrots: Can Language models Be Too Big?*, FAccT'21 (March 1, 2021) https://dl.acm.org/doi/10.1145/3442188.3445922.

The Office's interpretation of the operation of the model takes a very functional view. Their idea that because the models can create complex pictures like people do, they have autonomy as people, is rooted in the computational functionalist perspective, which argues that mental states - such as feelings, desires, and thoughts, are the same as computer processes in the brain.[208] It is the functional organization and structure of the brain that facilitates cognition, rather than the material substance.[209] What the brain is made up of is of less importance than what they do. As long as the other entities can do the same as the brain, it could be thought of as a brain.

Indeed, Putnam, a pioneer of this perspective and an early advocate for the computational view of the mind, argues that the mental states are fundamentally functional states.[210] To truly understand a human's belief requires people have insight into their functional organization. He says, "to know for certain that a human being possesses a particular belief, one must understand something about the functional organization of that human being."[211] Experiences such as pain are not just physical sensations; they signify a deeper, functional state of being harmed or injured.[212] As it is the function of an entity that determines its existential state, because GenAI can seemingly generate complex images, it must work autonomously and automatically as the people.

Another example to illustrate the misattribution of funcion to inner state is the Turing machine. Proposed by Alan Turing in the 1930s to solve the issue of computation, the Turing machine serves as a criterion to determine whether a machine can exhibit intelligent behavior equivalent to that of a human.[213] Its purpose is to make the machine indistinguishable from human activity.[214] The process involves a user engaging in a conversation with an unseen interlocutor, who could be either a human or a machine.[215] If the evaluator cannot reliably tell the machine from the human, the machine is considered to have passed the test. The foundational idea here is that if a machine can functionally replicate the cognitive processes of the human mind, it could be considered intelligent .

However, just because the machines can perform the same tasks as the human minds doesn't mean that it would be the same as human minds. What people have is much more than just performance. We aren't just our functional behaviors. We also have experiential qualities of sensations, feelings, perceptions, thoughts, and desires, or simply, "qualia."[216] This phenomenal, and sensational content of our experience cannot be replicated by mechanical processes.

---

[208] *See generally* Oron Shagrir, *The Rise and Fall of Computational Functionalism* 220-250 (Yemima Ben-Menahem ed., Cambridge University Press 2005).

[209] *Id.*

[210] *See Philosophy of Mind of Hilary Putname,* BRITANNICA, https://www.britannica.com/biography/Hilary-Putnam/Philosophy-of-mind (last visited March 3, 2024).

[211] *See* Oron Shagrir, *Hilary Putname and Computational Functionalism,* PHILOSOPHY OF MIND: THE KEY THINKERS (June 3, 2013) 147, 150 https://openscholar.huji.ac.il/sites/default/files/oronshagrir/files/putnam_and_computational_functionalism_chapter _8.pdf.

[212] *Id.* at 157

[213] *See* Yongjun Xu at al, *Artificial Intelligence: A Powerful Paradigm for Scientific Research,* THE INNOVATION, VOL.2, ISSUE 4, (Nov 28, 2021), https://www.sciencedirect.com/science/article/pii/S2666675821001041.

[214] *Id.*

[215] *Id.*

In conclusion, the Office's perspective that GenAI models, due to their ability to perform tasks similar to human artists, possess comparable levels of autonomy, conceptualization, and execution might stem from the computational functionalist approach in cognitive science. This approach argues that the mind operates through a set of functions – what it does defines what it is. However, this view is flawed because it disregards the subjective properties of human intelligence. GenAI, in spite of their impressive abilities to generate complex pictures, aren't capable of true understanding.

**b. Understanding the fundamental Gap between GenAI and Human creativity: Beyond Mere Symbol Manipulation**

GenAI fundamentally lacks the creative intent and adaptability intrinsic to human cognition. There's a significant gap between AI statistical manipulation and genuine human intelligence. The models' clever manipulations of tokens, patches, or symbols do not equate to genuine comprehension of the content; the artists' incorporating them in their workflow doesn't mean they lose control over the creative process.

Compared to the algorithmic models, the human mind is far more dynamic in its processing of symbols and concepts. Jerry Fodor, in his seminal work T*he Language of Thought*, introduced the concept of "mentalese," a hypothetical mental language in which thinking occurs. [217] According to Fodor, this internal language uses symbolic "words" to denote tangible, real-world objects.[218] These "words" are organized into "sentences," with their meanings emerging from the diverse combinations and arrangements of these symbols, much like a personal library filled with books in a language only comprehensible to the owner. [219] In this mental library, each "word" corresponds to a real-world item—for instance, "tree" symbolizes an actual tree, and "dog" represents an actual dog. These words are assembled into sentences within the mind, crafting mental images or ideas, such as a dog chasing a cat, from the combination of "dog," "chases," and "cat." [220] Based on the composition and arrangement of the words, the sentences could have different implications.

When we contrast the dynamic process of human cognition with that of GenAI, the limitations of AI become even more evident. Although in both cases, real-world objects are represented in a language that the mind or the model can understand, GenAI works merely through manipulations of symbols. It would be like John Searle's classic Chinese Room: a person inside a room receives Chinese characters and manipulates them intelligently according to a set of rules, despite having no understanding of Chinese. To an external observer unaware of the internal workings, it might seem as though the person inside the room is communicating fluently in Chinese, genuinely comprehending the language. However, the reality is that the individual is merely following syntactic rules without any real understanding.[221] Consequently, although a GenAI model can skillfully manipulate signs, by no means would this behavior suggest a conscious understanding of the meaning and context.

---

[216] *See generally* Ryota Kanai & Naotsugu Tsuchiya, *Qualia*, CURRENT BIOLOGY VOL. 22, ISSUE 10 (May 22, 2012), https://www.sciencedirect.com/science/article/pii/S096098221200320X#:~:text=The%20phenomenal%20aspect%20 of%20consciousness,%27%20or%20%27what%20kind%27.

[217] *See The Language of Thought Hypothesis*, INTERNET ENCYCLOPEDIA OF PHILOSOPHY, https://iep.utm.edu/lot-hypo/ (last visited March 3, 2024) (Suggesting in the section for combinatorial syntax and compositional semantics that thoughts occur in a formal mental language termed mentalese); *See also The Language of Thought Hypothesis*, STANFORD ENCYCLOPEDIA OF PHILOSOPHY (May 28, 2019), https://plato.stanford.edu/entries/language-thought/.

[218] *Id.*

[219] *Id.*

[220] *Id.*

## D. Exploring the Reluctance: Why the Copyright Office is Hesitant to Spend Serious Efforts to understand the mechanism of GenAI

The reluctance of the Copyright Office to thoroughly explore and understand the intricacies of text-to-image generators and its application in art creation may be inadvertently influenced by a common human tendency: a diminished inclination to engage deeply with what is perceived as less valuable. This perception, particularly when applied to computer-generated art, could stem from two factors: the tension between subjective and objective evaluations of art and the perceived absence of the performative element associated with computer-generated art. This section seeks to establish that 1) there's a prevalent skepticism towards works generated by machines, and 2) because computer art lacks the performative element and the perception of human involvement, people are less inclined to view it as valuable as art created by humans. As a result, the Copyright Office is less willing to spend serious efforts to understand the mechanism behind the technology. Of course, this isn't to claim that the Copyright Office is deliberately engaging in value assessment of AI-generated art before considering authorship. It simply means that the Office, like others, could be caught in an innate human bias. This bias might inadvertently affect their readiness to explore the technological intricacies of AI in art.

### a. Perceptions of AI-Generated art: Navigating the Dichotomy of Subjective Experience and Objective Standards of AI Art in Copyright Consideration

This section seeks to explore how subjective and objective evaluations of art influence the Copyright Office' decision to engage deeply with the mechanism of GenAI models. It argues that while objectively, a work of art may meet the aesthetic of beauty, subjectively, it fails to resonate on a deeper level with the audience, leading to Office's hesitancy to understand the technology.

It will have two parts: the first part discusses the dichotomy between objective and subjective evaluations of art. It proposes that this division is rooted deeply in the social philosophical and aesthetic traditions. The second part presents empirical evidence to support the notion that there exists a broad skepticism towards the artistic value of machine-generated works, which affects not only individual perceptions but also institutional attitude to them.

#### 1. Subjective vs. Objective Evaluation of Art.

Traditionally, art evaluation is divided into two distinct schools: subjective, which is based on personal preferences and experiences, and objective, which considers measurable aspects like symmetry, the relationship between elements, and the interplay of light and shadow.[222] These two perspectives may not always align; it's possible for someone to appreciate the objective beauty of a piece while subjectively disliking it or vice versa. In the case of AI-generated art, people might acknowledge its objective beauty but still reject it due to anthropocentric beliefs.

---

[221] *See generally*, MARGARET A BODEN, COMPUTER MODELS OF MIND 89-104 (John Heil ed., Cambridge University Press, 1988)

[222] *See* Christopher P Jones, *Subjectivity and Objectivity in Art*, MEDIUM (Dec 13, 2019) https://christopherpjones.medium.com/subjectivity-and-objectivity-in-art-cc41d55c76a5.

Subjective evaluation of art focuses on the idea that beauty is fluid and personal, varying from one individual to another based on their unique experiences and perceptions. This concept was first proposed by Enlightenment philosophers David Hume and Immanuel Kant.[223] Hume asserted that beauty is in the eye of the beholder; aesthetic appreciation is an inherently personal experience: "Beauty is no quality in things themselves: It exists merely in the mind which contemplates them; and each mind perceives a different beauty. One person may even perceive deformity, where another is sensible of beauty; and every individual ought to acquiesce in his own sentiment, without pretending to regulate those of others."[224]

For Hume, beauty is not an inherent quality but rather a perception that changes with each observer. This idea is echoed by Kant, who believes that aesthetic judgments are deeply personal, grounded in the emotional responses elicited by an artwork. He suggests: "the judgment of taste is therefore not a judgment of cognition, and is consequently not logical but aesthetical… through which there's a feeling in the subject as it is affected by the representation."[225]

The objective standard, however, means that beauty isn't personal. Instead, there exists an external standard of assessment against which beauty is evaluated. This school dates back to ancient philosophers like Plato and Aristotle, who argue that beauty is judged by metrics.[226] Plato, for example, conceptualizes an ideal Form of Beauty, implying that the beauty we perceive on earth is merely a reflection of this ultimate perfection.[227] In *The Republic*, he articulates that the physical world and all its beauty are but shadows of the true, eternal Forms that exist in a realm of perfect and unchangeable ideas, including the Form of Beauty itself.[228] Aristotle, in the meantime, perceives beauty as emanating from the natural attributes of symmetry and balance.[229] This viewpoint is discussed in the *Poetics* where he delves into the aesthetic foundations of art and literature, underscoring the significance of harmonious proportions.[230] His appreciation for these qualities is also implicit in *Metaphysics*, where discussions about the structure and essence of the universe echoes his understanding of beauty as inherently linked to the natural world's orderly and symmetrical arrangements.[231]

Therefore, even though AI-generated art may meet objective standards of aesthetic appeal, an individual's personal experiences can still lead to disapproval of the work. This aversion, combined with the complexity of the art's generative process, may cause the Office to rely on its biases without pursuing further investigation.

---

[223] *Beauty*, STANFORD ENCYCLOPEDIA OF PHILOSOPHY (Sep 4, 2012), https://plato.stanford.edu/entries/beauty/.

[224] *Id.* citing (Hume 1757, 136); *see also Hume's Aesthetics,* STANFORD ENCYCLOPEDIA OF PHILOSOPHY (Dec 17, 2003), *https://plato.stanford.edu/entries/hume-aesthetics/*

[225] Immanuel Kant, *Critique of Judgment,* Analytic of the Beautiful, §1.

[226] *Beauty*, STANFORD ENCYCLOPEDIA OF PHILOSOPHY (Sep 4, 2012), https://plato.stanford.edu/entries/beauty/.

[227] *See Plato's Aesthetics*, STANFORD ENCYCLOPEDIA OF PHILOSOPHY (Jun 27, 2008), https://plato.stanford.edu/entries/plato-aesthetics/

[228] *See* Plato, *Phaedrus* 250d-256b; *Symposium*, 210a - 212a, from the speeches by Socrates and Diotima.

[229] *See* Aristotle, *Metaphysics* Book XIII; *See also* Aristotle, *Poetics*, Ch. 7, discussing how sense of order and proportion is indispensable for creating a beautiful and aesthetically pleasing narrative.

[230] *See Aristotle's Aesthetics*, STANFORD ENCYCLOPEDIA OF PHILOSOPHY (Dec 3, 2021), https://plato.stanford.edu/entries/aristotle-aesthetics/.

[231] *See generally,* John S. Marshall, *Art and Aesthetic in Aristotle,* THE JOURNAL OF AESTHETICS AND ART CRITICISM, VOL. 12 NO. 2, 228-231 (Dec. 1953), https://www.jstor.org/stable/426876.

## 2. Empirical Evidence for Diminished Objective Value for AI art

Empirical evidence from psychological studies confirms the idea that people implicitly devalue the art when they know that the work is created by algorithm or computers. In a study conducted by Di Dio et al. to investigate the effect of artist identity on subjective and objective valuations, Di Dio et al. enrolled 139 Italian adults who have had no formal education in the arts.[232] They used a multifactorial repeated-measures design to study the interactions among conditions (blind, primed), aesthetic judgments (beauty, liking), and authorship (human, robot).[233] Their findings revealed that when participants recognized the piece was created by machines, the objective value they attributed to the work diminished; conversely, when the work was identified as produced by a human, the subjective value increased.[234] This means that when people realize that the work they are looking at is created by a machine, they attribute less artistic competence to it; they think that because the creator is a machine, it is expected to produce complex things. The machine will invoke less wonder and resonance in the audience compared to a work created by a person.

This suggestion that people are less likely to appreciate machine-generated work is consistent with other studies. Millet et al. designed 4 studies, involving 1708 participants to investigate the reaction towards AI-generated art and its correlation with individuals' beliefs about human nature.[235] The study showed that participants valued artwork less when they knew it was created by AI rather than by humans.[236] They perceived computer-generated pieces as less creative, leading to a reduced sense of awe and diminished sense of enjoyment in art.[237] For individuals who strongly believed that creativity was a uniquely human attribute, this effect was even more pronounced; they found AI-generated art less impactful.[238]

Therefore, despite the objective beauty of AI-generated works, people's anthropocentric beliefs may lead them to devalue these creations artistically. This unconscious bias against AI art may make the Office less willing to spend efforts to understand and investigate the mechanisms of technology, leading to the perception that it is automatic and random as claimed in the Guidance.

## b. How AI Art's Lacking the Performative Aspect Influences Copyright Office's Willingness to Investigate the Technology

In addition to the Office's potential to attribute less subjective value to AI-generated art, the perception that AI art lacks a performative element and is created with less efforts could make the Copyright Office biased against it even before the evaluation begins. This does not imply misjudgment on the part of the Copyright Office, but rather, it highlights common human biases that are inherent to our nature.

---

[232] *See* Di Dio et al, *Art Made by Artificial Intelligence: The Effect of Authorship on Aesthetic Judgments*, Psychology of Aesthetics, Creativity, and the Arts (2023) https://psycnet.apa.org/record/2023-98729-001.
[233] *Id.*
[234] *Id.*
[235] *See* Kobe Millet et al., *Defending Humankind: Anthropocentric Bias in the Appreciation of AI Art*, Computers in Human Behavior Vol 143 (June 2023), https://www.sciencedirect.com/science/article/pii/S0747563223000584#sec3.
[236] *Id.*
[237] *Id.*
[238] *Id.*

This section is divided into two parts. The first part addresses the perceived absence of performative elements in computer-generated art. It suggests that while human-created art often involves a visible display of skill, emotional depth, and unique artistic expression to emotionally engage audiences, these elements are frequently seen as missing in AI-generated art. This absence can hinder the formation of a deep connection between the artist and the audience. The second part explores the psychological aspects, drawing on studies in psychology to argue that the perceived ease in creating computer art may contribute to a bias against it, as people tend to not value the things they spend less effort on. This bias may, in turn, reduce the willingness to engage in and understand the technology and artistic merit behind AI-generated works.

**1. Computer Art lacks the Performative Aspect of Art**

Computer art is often perceived as merely a final product, missing the performative essence that characterizes human artistic creative process. As a result, the Copyright Office may subconsciously not treat it on the same level as the human created art.

A work of art is a performance; it is more than a static image for passive viewing. It's an interactive experience that creates a dynamic space for engagement between the creator and the audience.[239] Art made by humans incorporates twelve essential elements, such as technical skill, unique style, the capacity to engage and surprise, authentic reality depiction, personal expression, emotional depth, intellectual stimulation, and a collective imaginative experience shared between creators and viewers.[240] These components emphasize the artist's role in engaging the audience. The idea would be that - when viewers observe a piece of art, they see an artist using their skills and resources to manifest their creative vision. Every brushstroke, chisel mark, color choice, and the placement of elements contribute to the work's authentic expressiveness. Although the final artwork stands alone and demands attention, the human influence within it is unmistakable. This human aspect is vital for a deep and layered understanding of the art.

In contrast, computer-generated art often misses the personal narrative that defines human-created works. An algorithm doesn't embody artistic performance in the same way a human artist does. When individuals employ algorithmic models for art creation, they become obscured behind the mathematical processes. The viewers see mathematical formulas, rather than a live human that expresses themselves. Therefore, these algorithms strip away the human essence, presenting the artwork as a product of computational probabilities rather than deliberate, tactile manipulations.

As a result, when comparing computer-generated art to human-created art, viewers may not engage with them on the same level. While art created by humans could foster a deeper emotional and personal connection, art produced by machines can lack the performative aspect that draws people into an emotional bond with the work. This absence of interaction between the observer and the artwork, and the inherent performance in the work could lead to a potential bias against it by the Copyright Office even before it embarks on the authorship inquiry.

---

[239] Dorothea von Hantelmann, THE EXPERIENTIAL TURN, https://walkerart.org/collections/publications/performativity/experiential-turn/ (last visited March 3, 2024)
[240] *See* John Valentine, *A Note on Denis Dutton's Concept of Art*, FLORIDA PHILOSOPHICAL REV, VOL. XIV, ISSUE 1 (2014) https://cah.ucf.edu/fpr/article/a-note-on-denis-duttons-concept-of-art/ (The other elements include intrinsic enjoyment, constructive feedback, an extraordinary focus, recognition by art institutions, active involvement)

**2. Effort Perception of AI art: Influencing the Copyright Office's Willingness to Investigate**

Another factor that might contribute to the Office's bias against AI-generated art and their reluctance to thoroughly investigate the mechanism of technology might be the belief that art produced by computers requires little effort. This perception might lead to the devaluation of art created by algorithms. The Office might suffer from unconscious bias against it because as a principle, people are generally less inclined to invest significant effort in understanding things they deem low in value. This bias could affect the Office's willingness to spend efforts in investigating this technology right from the start.

Numerous research have demonstrated that people tend to value things they spend more effort on. In 1959, Aronson and Mills, in their classic experiment, explored the idea of effort justification within the framework of cognitive dissonance theory.[241] They found that individuals who underwent a more strenuous initiation process to join a group tended to value the group more highly than those who went through a less demanding process. [242] The study involved 63 female college students as participants.[243] They were divided into three groups: severe initiation, mild initiation, and control.[244]

In the severe initiation condition, participants were required to read embarrassing materials aloud before joining the group; in the Mild initiation condition, the material was less embarrassing; and in the Control condition, no material was read before joining.[245] All participants listened to the same recorded group discussion and then evaluated it.  Aronson and Mills found that participants in the Severe initiation condition showed a significantly higher level of liking for both the discussion and the participants of the group compared to those in the Mild and Control conditions.[246] The difference in ratings between the Severe and Control conditions was particularly pronounced, reaching a 0.01% level of significance, while the difference between the Severe and Mild conditions reached the 0.05 level of significance.[247] These findings suggest that the increased effort and discomfort experienced during a severe initiation led to a greater appreciation for the group.

Daryl Bem's self-perception theory also sheds light on a more subtle aspect of how people come to understand their own feelings, attitudes, and internal states through the external environment and conditions. According to Bem, this understanding often arises from observing one's own actions and the contexts in which these actions unfold, rather than from internal cognitive dissonance.[248] For example, individuals might gauge their preference for something based on the amount of time and effort they've put into it, rather than through introspection. If a person invests significant effort towards a goal, they are likely to perceive that goal as valuable, enhancing their appreciation of the result.

Indeed, the principle that increased effort leads to greater appreciation is supported by various other studies. Festinger and colleagues, for instance, argued that people naturally strive for consistency between the effort they expend and the value they attribute to their achievements.[249] When individuals notice a misalignment between their actions and beliefs—especially after undertaking challenging tasks—they tend to adjust their perceptions to match their efforts, thereby alleviating any tension or dissonance they might feel.[250] Similarly, research by Gerrard and Mathewson has shown that when people engage in demanding activities, their increased effort results in a higher valuation of the outcomes.[251] Wicklund and Brehm also find that when individuals view their choices as self-determined, particularly after investing effort, this sense of autonomy boosts the value they place on the results of those choices.[252] Conversely, when the effort seems minimal, the resulting work is often less appreciated. This dynamic becomes particularly relevant in the context of AI-generated art. People might perceive it as requiring less effort compared to traditional human-created art, leading to an unconscious devaluation of the AI art.

---

[241] *See* Elliot Aronson & Judson Mills, *The Effect of Severity of Initiation on Liking for a Group*. The Journal of Abnormal and Social Psychology (1959), 177–181. https://psycnet.apa.org/doiLanding?doi=10.1037%2Fh0047195.
[242] *Id.*
[243] *Id.*
[244] *Id.*
[245] *Id.*
[246] *Id.*
[247] *Id.*
[248] *See* Daryl J. Bem, *Self-Perception Theory*, Advances in Experimental Social Psychology, Vol. 6 (1972) 1-62. https://www.sciencedirect.com/science/article/abs/pii/S0065260108600246

Applying these findings to AI-generated works – when people are presented with two pieces of art, one created by a human and the other by a machine, the audience tends to value the human-created work more. This is because the viewers can recognize the efforts that go into the composition of the painting, leading them to attribute more value to it. Conversely, when they view art produced by a computer, they perceive it as requiring minimal effort, and thus, they assign it less value. This perception of reduced engagement in the creative process predisposes people against art created by computers.

Of course, this isn't to imply that the Copyright Office engages in simultaneous value judgment of a work of art when determining its authorship. It simply means that the Copyright Office, just like anyone else, carries their own biases and perspectives when faced with new developments, such as AI-generated art. These biases can affect their actions, including the level of effort they are prepared to put into to understand the technology behind AI art. This limited understanding could unintentionally lead to a misinterpretation of the technology and its capabilities, influencing their copyright-related decisions.

## III. The Office Misunderstands the Dynamic Nature Between the Models and Their Users.

In the Guidance, the Copyright Office determines that the model users are not eligible for copyright protection because they do not possess "ultimate creative control over how systems interpret prompts and generate material."[253] It compares the prompts given to the models by the users to "instructions to a commissioned artist," with the system automatically and autonomously filling in the details such as style, composition, themes and angles. As this section shows below, this interpretation is incorrect. It's based on a misunderstanding of how the users actually use the tools – just because artists integrate the models in their creation process, it doesn't mean that they are surrendering control over the final output. At each stage, the model users could always step back to assess whether what is being depicted reflects the image they have in mind.

---

[249] *See generally* LEON FESTINGER, A THEORY OF COGNITIVE DISSONANCE (Stanford University Press, 1962); *What is Cognitive Dissonance*, MEDICALNEWSTODAY, https://www.medicalnewstoday.com/articles/326738 (last visited March 3, 2024)

[250] *Id.*

[251] *See* Harold B Gerard & Grover C Mathewson, *The Effects of Severity of Initiation on Liking for a Group: A Replication*, JOURNAL OF EXPERIMENTAL SOCIAL PSYCHOLOGY, VOL. 2, ISSUE 3 (1966), https://www.sciencedirect.com/science/article/pii/0022103166900849.

[252] *See generally* ROBERT A. WICKLUND & JACK W. BREHM, PERSPECTIVES ON COGNITIVE DISSONANCE (Erlbaum, 1976), https://psycnet.apa.org/record/1977-04655-000

[253] Copyright Registration Guidance: Works Containing Material Generated by Artificial Intelligence, 88 FED. REG. 16,190 (Mar. 16, 2023) (to be codified at 37 C.F.R. § 202).

This section will be divided into three subsections. The first subsection presents three case studies: a traditional artist who incorporated GenAI into his workflow, Suzanne Treister, and Jason Allen. These examples illustrate that, regardless of their varying degrees of experience in the professional art world, when they use GenAI, all of them adjust, refine, assess, and adopt the changes at each step of the creation; they are in control of the creation.

After establishing that the artists don't lose control over the creative process, the second subsection focuses on refuting the argument that the prompts given to the generative models are like instructions given to commissioned artists. I will use the example of Caravaggio's commission for *The Conversion of St. Paul* to show that the model users are far more engaged in the creation process than the clients of the commissioned artists. When they don't like the generated images, they change the elements themselves, rather than simply requiring the commissioned artists to start over.

In the third subsection, I will reapply the requirement of originality in *Feist* to emphasize that the significant role of the model users' input and creative engagement make them qualify for copyright protection. The element of "noise" in the models' training process doesn't take away the model users' control in the process. The users are in command of the process as much as Tim Knowles and Jackson Pollock are in their creative efforts.

## A. Artists Have Authorial Control Over the Final Output Because They Constantly Adjust, Refine, and Adopt the Changes at Each Step of the Creation. They Maintain Vigilance Over the Entire Process.

In the Guidance, the Office provides a fairly static and fixed form model of interaction between the users and the GenAI models. It argues that the users don't have ultimate creative control over the systems; it is the machine that ultimately determines how those instructions are implemented even though people provide the prompts.

As this section shows, this is an incorrect understanding of the nature of interaction between the users and the models. Typically, when the users integrate models in their creative process, they don't just become passive users that adopt and confirm whatever the model generates. Instead, they actively engage with this creation process, use the models as tools for inspiration and adjustment, and change the output iteratively. To say that the generative models strip people of their authorial control over the output is similar to suggesting that for Jackson Pollock who drips paint on canvas, because he doesn't control gravity, he doesn't exert control over his paintings.

To illustrate these points, I'll focus on three primary examples in this section using thematic analysis. This method helps identify recurring themes within the narratives. The first example comes from a YouTube video where a traditional artist shares his experience of using Stable Diffusion. Although he doesn't specifically mention why he chooses Stable Diffusion for his creative process, he emphasizes that the initial concept of the artwork originates from him. He expresses concern that starting with generative models might compromise the originality of his work. Unlike Jason Allen, who is significantly impressed by the generative capabilities of such models, this artist also develops his own models and stresses the importance of selecting the right samplers for training. Furthermore, in addition to Stable Diffusion, he also uses other tools such as Photoshop and Adobe, to achieve his desired results.

The second example features a clip from Art Basel, where contemporary British artist Suzanne Treister talks about her method of refining prompts and choosing outputs. The discussion highlights her view that these models are simply tools, not collaborators. Rather than integrating these tools into her artistic process like the previous artist, Treister explicitly refuses to use the generated images in her final works. Her interest lies primarily in probing, examining, and investigating the technology for its own sake. She rejects the idea that these GenAI models have autonomy and creative agency.

Jason Allen is the third example. He is the amateur artist who has crafted thousands of prompts for the Space Opera series. As the only non-professional in this comparison group, and an "outsider" to the traditional art world, Allen embraces the technology willingly. He admits to using the technology for brainstorming ideas for projects for his company; he also mentions his motivation to participate in the state fair - to be part of the bigger conversation about art, technology, and society.

All three examples show: 1) the more seasoned the artist in the art world, the less they tend to be dazzled by this technology, 2) this technology prompts a reexamination of our relationship with art, its impact on us, and its implications for society, and 3) regardless of the level of engagement, none of the artists mentioned here relies on a single iteration from these tools. All of them go through an extensive process of iteration, refinement, selection, editing, evaluation, and confirmation, contradicting the Copyright Office's suggestion in the Guidance that the technology operates automatically and randomly without human intervention.

### a. A Traditional Artist who Uses Stable Diffusion

The example below explores a traditional artist's experience incorporating Stable Diffusion into his workflow. It uses the artist's original, personal narrative to show 1) when he uses the technology, he perceives himself as the one directing the models during the image creation process, rather than being led by the models' outputs; 2) He views this technology as a gateway for individuals who may not possess the same level of experience or skill; 3). In using the technology, he actively iterates, modifies, and refines the prompts to achieve the desired components of the image. 4) He also uses additional tools, such as Photoshop to edit the picture. As a result, the final image reflects his creativity, personality, and conception.

The fact that the artist remained in control of the creation process when he used Stable Diffusion was obvious in his narrative. In the beginning, he mentioned his reason for portraying himself as the subject. He said, "*Stelfie is a very funny and very clumsy dude. He time travels and has the most incredible adventures. And he is sort of an alter ego myself, although physically we are completely different.*"[254] The reason he choose this subject was because he felt personally connected to it.

---

[254] Vox, *An AI Artist Explains His Workflow*, YOUTUBE (May 2, 2023), https://www.youtube.com/watch?v=K0ldxCh3cnI&t=126s.

After choosing the subject, he began conceptualizing the image by "*starting with, you know, drawing a sketch.*"[255] He didn't just write prompts to describe the picture, as although "*Stable Diffusion and the other diffusion models around are extremely good,*" they were also "*extremely cheeky.*"[256] He was worried that, "*It was very easy for them to drive you away from the original idea that you had.*"[257] So it was important to have your own original idea to work with.

Once he had a clear vision for his drawing, he experimented with "*a bunch of random prompts... just to see if [he] could find at least a good initial pose.*"[258] He found himself revising these prompts several times, explaining, "*[I] couldn't find the pose for Stelfie on the right. So, what I had done is that I moved to Photoshop...and I recreated the pose myself.*"

In addition to creating the initial pose himself, he tailored each body part using a variety of applications and tools. For example, for the face, he used a specialized model trained on Stelfie's distinctive features, "*I created Stelfie in 3D and captured numerous snapshots from various angles to train the model,*" he explained. But "*When it came to faces, it was very difficult to achieve...good results....So I've asked Stable Diffusion to make a face that would look like Muhammad Ali. And then in Photoshop, I warped all these traits. So I make the nose — larger or thinner. And the jaw… the eyes.*"[259] Eventually, he "*changed everything manually.*"[260]

For the pose, he aimed for a softer, less athletic build, because "*[he] wanted Stelfie to be like a bit fluffy in terms of not super fit, without a six pack. And probably here I sort of found… the belly that I was happy with. But clearly it was not realistic enough yet. And then I used the result and I modified the result in Photoshop.*"[261] This careful process of adjustment involved extensive modification—cropping, cutting, pasting, warping, and manual painting to perfect the arms, eyes, skin tone, and exposure.

Depicting hands presented a unique challenge, leading him to personally create 50% of them – "*Half of the hands in Stelfie's images were actually my own. And that was because it has always been extremely challenging to reproduce hands. So what I was doing is that, you know, if I needed the hand in some position I would take a picture of my hand...and then I would clean it up and paste on top.*"[262]

In addition to generative AI models, the artist leveraged various other editing tools to refine his work. He mentioned using ControlNet, an extension that significantly streamlined the process: "*If I needed to recreate the same pose I did two months ago, it would now take me about 15 minutes.*"[263] Throughout his creative process, he emphasized the importance of choosing the right samplers to achieve lifelike details, especially for textures like skin. "*Because the sampler was very important in terms of realism and details. So if you're trying to replicate skin… Euler was very synthetic, very fake. But DPM, for example, was working great on that. There were many parameters that were extremely important,*" he explained.[264]

The artist described a dynamic workflow, in which various parts of the artwork were fashioned using separate tools. "I looked back a lot between Stable Diffusion and Photoshop. So let's say that's out of 100%. 50% was done with Stable Diffusion about 40% in Photoshop and about 10% in Procreate."[265] The image wasn't the result of just one tool.

---

255 *Id.*
256 *Id.*
257 *Id.*
258 *Id.*
259 *Id.*
260 *Id.*
261 *Id.*
262 *Id.*
263 *Id.*
264 *Id.*

In addition to using various applications to edit specific body parts, he also fine-tuned his use of Stable Diffusion, carefully navigating through its multiple settings. He discussed the process of determining the number of iterative steps required, explaining, "So steps were how many times you were telling Stable Diffusion to work on your prompt. You could choose a very low number or a very high number. Many options."

Within this iterative framework of Sable Diffusion, he also highlighted the significance of "inpaint" and "outpaint" functions. "Inpainting allowed you to modify specific parts of the image, directing the AI to focus changes on selected areas. Outpainting, conversely, expands the canvas by having the AI extrapolate and add elements beyond the existing boundaries, based on the context of what's already depicted," he described.[266] This approach enabled him to adapt the tool's functionality to meet the unique demands of each aspect of his work, ensuring that the final product was not the result of a single, automated process but a series of deliberate, creative decisions.

When he ran into trouble revisiting and refining minor details, Stable Diffusion proved particularly helpful. He shared, "when I was about halfway through and encountered issues, I knew I could turn to Stable Diffusion for assistance. It was particularly useful for making precise adjustments, whether it was smoothing out edges, enhancing the lighting, or improving the texture of the skin. This level of detail was crucial for the overall quality of the artwork."[267] This flexibility to iteratively refine the work made sure that even the smallest elements met his exacting standards.

In his concluding thoughts, he emphasized the primacy of the artist's vision over the capabilities of the machine. He stated, "I felt you had to drive the machine, not the other way around. And just to prove how important was the artist's part in the overall process creation."[268] He viewed the creative process as a collaborative endeavor with AI, leveraging his two decades of experience as a traditional artist: "Because of my background in traditional canvas painting, I don't feel intimidated by AI,"[269] he explained. This technology was "an opportunity for a new generation of artists to explore a novel domain of art, distinct from existing digital art forms, and to discover fresh avenues for creativity."[270]

In conclusion, it was clear that he exercised complete control over every stage of the creative process—from the initial vision and conceptualization to creating the initial pose and adding additional ones; all these were purely his own ideas. The integration of technology didn't diminish his control; instead, it enhanced the efficiency and convenience of the creative process.

### b. Suzanne Treister

At the Art Basel 2023, Suzanne Treister, a pioneer in the new media field since the 1990s, shared her insights on Generative AI during a panel discussion. The session was aimed to explore the impact of recent advancements in AI image-generation technologies like Midjourney and DALL·E 2 on artistic expression and public perception. The discussion focused on how these developments influenced our understanding of creativity and agency; it questioned the extent of the collaboration between humans and AI in art creation; and the implications for the art world as these technologies grow more sophisticated. This section focuses on Treister's views and her engagement with these technologies. Unlike the artist mentioned in the last section, she doesn't incorporate he AI-generated elements into her workflow. Instead, she's deeply interested in understanding the technology, preferring to analyze its impact and raise questions about its effects on society.

---

[265] *Id.*
[266] *Id.*
[267] *Id.*
[268] *Id.*
[269] *Id.*
[270] *Id.*

During the discussion, Triester points out that she doesn't think of the models as independent and autonomous entities. She doesn't believe that they have agencies. As she mentions, "*This idea of like that, it felt like the machine had agency...I think it's wrong… Machines have no way of manifesting the works themselves in a physical world.*"[271] As it ultimately still depends on the human to bring out the creations in art, creativity belongs to the people; without people, machines wouldn't be able to produce anything.

The GenAI models, for Triester, are mere tools. Despite their abilities to produce complex images, Triester wouldn't appreciate them as more than technologies. She suggests that she "*currently doesn't have plans to carry out any of these machine intelligence generated works*."[272] She wouldn't take the machine-generated images too seriously. For her, the works "*[aren't] something that I want to use, [and when she does use the technologies,] it is to see what it would show me and what it would present.*"[273] She is interested in understanding and exploring the technology as an object, rather than treating it as a subject with autonomy.

Indeed, she tries repeatedly to get the machines to confirm its technological status. She says that "*if I asked it something I would ask it probably 10 times the same question and then refine it down and then I want to get it to acknowledge that it is machine intelligence.*"[274] She also says that "*sometimes you'd ask [the machine] something and they'd say I cannot answer that because I'm a machine intelligence.*"[275] In such situations, she would still go on insisting: "*I would go, well, how does it feel to be a machine intelligent and be thinking about this question of run around.*"[276] Eventually, the machine would generate "*something quite interesting although it's still machine intelligence.*"[277] The fact that the models aren't artists is very obvious for her: "*For you there was a definite distinction, you didn't have this whole uncanny valley phenomena you didn't feel like you're talking to a fellow artist.*"[278] She does highlight that this introduction of this technology encourages us to reconsider our relationship with art. In today's society, Triester argues, it becomes ever more important to question the implication of art to people and to society: "*What does the audience seek from art?... What draws them to art?...What does the general public expect from the art world? What are artists seeking from the art world? What are collectors looking for in art? And what about curators and museums—what do they seek from art?*"[279] And the existential question for art: "*Why is art necessary for the audience? What does art provide?*"[280]

---

[271] *See* Art Basel, *Conversations | Co-Creating with AI: The Artist's View*, YOUTUBE (June 17, 2023) https://www.youtube.com/watch?v=LjvM9lHP3dw.
[272] *Id.*
[273] *Id.*
[274] *Id.*
[275] *Id.*
[276] *Id.*
[277] *Id.*
[278] *Id.*
[279] *Id.*
[280] *Id.*

**c. Jason Allen**

Jason Allen is the only non-professional artist in this comparison group. Contrary to the Copyright Office's suggestion in their rejection of Allen's second registration request—that his work consists merely of choosing prompts—Allen's creative process is significantly more complex, marked by ongoing adjustments and refinements.

Unlike the two professional artists mentioned earlier, Jason Allen comes from a technology background. Before the introduction of AI-generated art, his artistic experience primarily centered around anime art. He admitted that he "*liked anime art manga… and was pretty good at [drawing it].*"[281] Indeed, he introduced himself as a "*game designer and art director for incarnate games incorporated.*"[282] He was "*the president and ceo…of the company.*"[283] They created lots of games  – "*board games, tabletop games, card games,  dice games.*"[284] He worked in the intersection of tech, games, and art.

The reason he started using the technology was because he became amazed at the generative capabilities of the models and he wanted to play with it. He admitted that he was a technology guy. Although he didn't "*have a computer science degree and [he] didn't code anymore… this stuff still interested [him].*"[285] So he just started playing with it. He "*started playing around with night café starry AI…and wombo dreams on the cell phone;*" he also "*started doing images of the second coming of christ.*"[286] And he "*was shocked at some of the images that [he[ was seeing.*" He thought they were "*unbelievable.*"[287] Then, a guy on twitter invited him to Midjourney. And now he was "*just like a kid in a candy store and running all kinds of tests because there was a certain method to the madness you couldn't just say.*"[288]

He mentioned that he entered the competition at the state fair for two reasons - first, he would like this experience to be part social experiment; second, he would like to be part of the bigger conversation about the place of tech in society. He mentioned that "*I was bringing out you know, my full ideas onto the discord mid journey and I started realizing…some of these were really good. I wanted other people to experience this. I realized the state fairs were coming up and maybe it would be cool to enter this into the contest, and just see how it went. It would at least further the debate.*"[289] His idea was that if he did well, he could "*bring that debate out of discord and into the general public. And so to speak, just be part of the larger conversation.*"[290] He was here to help people understand that the technology wasn't "*as terrifying as as you think and the ethics of it shouldn't be pointed at you. [It] shouldn't be pointing at the people who are choosing to use this tool as the method or outlet for their work and also you shouldn't be demonizing the technology.*"[291] He hoped that, through him, once people know more about the tech, they could form their own opinions about this disruptive tech.

---

281 *See* Disruption Theory, *How Jason Allen Used AI to Win Art Prize*, YouTube (Sep 9, 2022) https://www.youtube.com/watch?v=yAQz7NO2E_U&t=30s.
282 *Id.*
283 *Id.*
284 *Id.*
285 *Id.*
286 *Id.*
287 *Id.*
288 *Id.*
289 *Id.*
290 *Id.*
291 *Id.*

As to how he used the technology, he didn't object to using it for ideation and conceptualization. He admitted that his company already used the tech for ideation of the project they were working on. He said that "*we had already used it for ideation. There was a particular design element that dylan pierpont, the lead artist to this project, that we were working on now. And we looked at some specific design elements that I created with mid journey that we used to incorporate into the cover of our new title, which was still to be announced.*"[292] He didn't find it absolutely necessary to leave the step of conceptualization to himself.

However, just because he used it for ideation didn't mean that he wholeheartedly accepted the generated images without modification. He did confirm that there were lots of elements that AI couldn't get right. He mentioned, "*you might not get the result that you were looking for specifically. Not yet anyway. AI was still evolving. The technology was still advancing. Usually when you were an art director lead on a project, there was a very specific outcome that you wanted to have about how things were placed. elements that were incorporated; the angles that were used, the lighting that was used, how it was going to be perceived from a distance and up close…there was just so many things that you had to consider that AI might actually not be able to do right off the bat.*"[293]

Indeed, when describing his creation process for the Space Opera Theater, Allen described it as an extensive process involving long hours of adjusting prompts, fine-tuning themes, styles, and elements. He mentioned that he "*changed the text prompts with every creation.*"[294] He also experimented with new settings, scenarios, and effects. For inspiration, he started with a mental image of "*a woman in a Victorian frilly dress, wearing a space helmet,*" because he had never seen it. Then, for the scenes, he kept fine-tuning the prompts. He said that he used "*sets to make an epic scene, like out of the dream.*"[295] He also admitted to having spent 80 hours making more than 900 iterations of the art, adding words like "*opulent*" and "*lavish*" to fine tune its tone and feel.[296]

Unlike the previous two artists who utilized various tools to alter scenes, Allen placed significant emphasis on prompts. In contrast to the first artist, who developed their own model and curated samplers for model training, Allen's approach for the Space Opera Theater involved developing a foundational prompt that remains unchanged. He stated – "*so space opera theater has a core foundational string that was the main prompt that hadn't changed since I started.*"[297] Beyond this, he extensively modified elements like "*the composition, the aspect ratio, the lighting,*" and engaged in post-processing, focusing on, "*a lot of different little elements that [he] would focus that I would change as I generated different iterations.*"[298] These adjustments were time-consuming.

When merely adjusting and refining prompts didn't yield the desired results, Allen turned to additional tools, such as Adobe Photoshop, to remove visual artifacts from selected images. For example, he added dark, wavy hair to a central figure that was missing its head. To further improve the photos' quality and sharpness, he used another AI tool, Gigapixel AI. He then printed the images on canvas. Then, he transported these canvases to the state fair.[299] The artwork he submitted was the product of

---

292 *Id.*
293 *Id.*
294 *Id.*
295 *Id.*
296 *Id.*
297 *Id.*
298 *Id.*
299 *Id.*

thousands of iterations, involving extensive use of Photoshop for editing, selection, and rearrangement

He compared using the model to describing scenes in movie scripts. He said that crafting prompt wasn't "*like talking to a person or telling a story.*"[300] It was not about direct communication to an individual. Rather, it was "*a little bit like when we discussed scenarios and scenes like a director to a movie.*"[301] It was only when the prompts were explained well that the pictures "*really started to take shape and become more than what [he] was seeing before.*"[302] The more detailed the descriptions, the better quality the pictures had.

He believed one could use the tech to address the writer's block. For example, imagine a setting where an artist was looking for a specific scene. She didn't know "*where [she] wanted the building to be and [she] couldn't figure out the composition.*"[303] She could "*come over here, use AI … just go through a few iterations.*"[304] And perhaps she would discover "*the [exact] composition [she] was looking for.*"[305] The purpose of using AI wasn't to replace the artist, but to reposition them as the heart of the idea generation and creation.

In conclusion, although Triester is the only one among the three who is interested in this technology as is, both the traditional artist and Jason Allen have leveraged it extensively for refinement and adjustment. Allen's use of the tool diverges significantly from the process described in the copyright office's registration denial. He doesn't merely tweak prompts and choose from thousands of generated images; he begins with his own ideation and conceptualization. He then employs the tool to modify style, elements, and lighting and shadows, further refining his work with Photoshop. This multifaceted approach shows his control over the creative process.

## B. Greater Control and Engagement: How Model Users Redefine the Creative Process Compared to Traditional Commissioned Artistry

In the Guidance, the Copyright Office draws a parallel between generative models and commissioned artists, suggesting that prompts provided by the model users serve as instructions similar to those given to a commissioned artist, guiding the final output.[306] However, this analogy is somewhat simplistic. In situations where the users provide a brief textual description and accept the model's output as is, this comparison might seem apt. However, this comparison doesn't hold up in more complex scenarios where artists use these tools for detailed and nuanced creations. The prompts are one of the many means through which the users refine their personal vision and conceptual framework. They are much more intimately engaged with the creative process than clients of commissioned work. This distinction becomes clear when considering historical commissions, such as Caravaggio's assignment by Monsignor Tiberio Cerasi, Treasurer-General to the Apostolic Chamber, to paint *The Conversion of Paul*, where the artist's interpretation played a significant role.

Consider the contract Cerasi engaged Caravaggio. On Sep 24, 1600, the document reads:

---

[300] *Id.*

[301] *Id.*

[302] *Id.*

[303] *Id.*

[304] *Id.*

[305] *Id.*

[306] Copyright Registration Guidance: Works Containing Material Generated by Artificial Intelligence, 88 Fed. Reg. 16,190 (Mar. 16, 2023) (to be codified at 37 C.F.R. § 202).

> Michael Angelo Merisi da Caravaggio…outstanding painter of the city, contracts with Tiberio Cerasi to paint two pictures on cypress wood, each with a length of ten Roman palmi and a width of eight, representing the Conversion of St Paul and the Martyrdom of St Peter, for delivery within eight months, with all figures, persons, and ornaments which seem to fit to the painter, to the satisfaction of his Lordship.
>
> The painter shall also be obliged to submit specimen and designs of the figures and other objects with which according to his inventions and genius he intends to beautify the said mystery and martyrdom. This promises the said painter has made for an honorarium and price of 400 scudi in cash… [having received] 50 scudi in the form of a money order directed to the Most Illustrious Vincenzo Giustiniani … for all this the parties have pledged themselves… they have renounced to the right of appeal, in perfect consent and have taken their oaths respectively…[307]

Caravaggio is required by the contract to paint 2 pictures for Tiberio Cerasi. The requirements of the two pictures are: long ten Roman Palmi, wide 8 palmi; painted on cypress wood; delivered in eight months; representing the Conversion of St Paul and the Martyrdom of St Peter. The rest of the elements such as interpretation of stories, and ratio of light and shadow are left to Caravaggio's personal choice. As read in the contract, all "specimen and designs of the figures and other objects" are to be painted "according to his inventions and genius." "All figures, persons, and ornaments" shall be depicted in ways that "seem fit to the painter." The painter could "intend to beautify the said mystery and martyrdom." As long as the outcome is "to the satisfaction of his Lordship," Caravaggio is expected to paint the picture in his characteristic style of intense emotions.

Compare the client's instructions to the model to Jason Allen's process of creating the Space Opera Theater series, it is apparent that Jason Allen is far more engaged in the creation process and maintains a significantly higher level of control than Cerasi. Unlike Cerasi who simply mentions the theme of the Work, Allen doesn't just ask the models to give him representations of the themes he wants to depict. Instead, he goes through an elaborate process of starting with a simple mental image - "a woman in a Victorian frilly dress, wearing a space helmet." Then, he keeps fine-tuning the prompts, using tests to nail down an epic scene. Eventually, he spends over 80 hours making more than 900 iterations of art.[308] He also uses Photoshop for better refinement.[309] One could never imagine a client having this level of engagement in the painting process after already contracting a commissioned artist.

Another aspect where the generative model diverges from the traditional commissioned artist is in their response to rejections by clients. When a commissioned artist creates a piece that fails to meet the client's expectations, the client is free to reject the artwork outright, thus requiring the artist to begin anew, often without further instructions. In contrast, with generative models, should the outcome not meet the user's approval, the burden falls on the user to refine and adjust their input descriptions. This iterative process, aimed at guiding the model towards a more accurate understanding of the user's vision, demands a greater level of engagement and effort compared to the more straightforward artist-client interaction.

---

[307] *See* Andrew Graham-Dixon, Caravaggio: A Life Sacred And Profane 211 (W.W. Norton & Company, 2012)
[308] *See* Drew Harwell, *He Used AI to win a Fine-Arts Competition. Was it Cheating?*, Washington Post (September 2, 2022, 11:08 am) https://www.washingtonpost.com/technology/2022/09/02/midjourney-artificial-intelligence-state-fair-colorado/
[309] *Id.*

For example, consider once again Caravaggio's painting *The Conversion of Paul*. After Caravaggio delivers the painting, Cerasi rejects them. The composition is cluttered. As Paul squirms on the ground, covering his eyes from the dazzling celestial vision, his horse rears up and foams at the mouth.[310] The saint's elderly attendant, clutching a shield adorned with a crescent moon and donning an elaborate plumed helmet, looks like a confused spear-carrier in a comic opera.[311] Caravagio has no choice but to start over.

The second time isn't easy. Caravaggio is visibly constrained by Michalangelo's influence:

> During his early struggles with the Cerasi Chapel Commission, Caravaggio was handicapped by an apparent inability to get away from the famous prototype of Michalangelo's restless and turbulent *Conversion of St Paul* in the Pauline Chapel. The rearing horse and reeling saint, the figure of Christ descending from the heavens, arm outstretched - he borrowed and adapted all these elements from Michelangelo's far larger and more densely populated painting, as if he were setting out to create a condensed version of the earlier work. It was only when Cerasi rejected the painting out of hand that Caravaggio reconsidered and found a diametrically different solution.[312]

Cerasi doesn't care what Caravaggio is struggling with when he paints the picture. For Cerasi, as long as the painting isn't satisfactory, he can ask Caravaggio to start over. The fact that Caravaggio borrows his style from Michelangelo doesn't matter much to Cerasi.

As Andrew-Dixon notes, the second time Caravaggio paints,

> Gone are the creakingly theatrical figures of Christ and the angel, replaced by a spectral radiance that is the light of God. There is no noise, no clamor, no comedy of misapprehension here - just simple ignorance contrasted with miraculous divine illumination, an irresistible tide of light that floods the saint and changes him forever.[313]

Had Caravaggio not re-read the Acts of Apostles, he might not be able to have this revelation that helped him create this painting that goes down history as a masterpiece; had he simply followed the conventional styles of drama when telling the story of St. Paul, his picture might not be characterized by his own style. It is his own interpretation of the story, and his personal choice of re-reading the *Acts* that helped him create the painting. His client, Cerasi, doesn't provide any further instructions or guidance. The task falls solely on Caravaggio to re-characterize and finish the painting.

Compare this to the creation process of the model users. When the generative models generate an outcome that doesn't meet the expectation of the users, it is the users, rather than the models, that must engage to change the output to make it align with their visions. For example, when Midjourney generates versions of space opera theater that isn't to the satisfaction of Jason Allen, it is Allen, rather than the models that must spend time, effort, and energy to edit, adjust, and confirm the output. The task is much more than just editing prompts and selecting the final outcome, as one could never expect the generative models to find new inspirations voluntarily. For the models to work, the users must be engaged in image creation at every step.

---

[310] *See* ANDREW GRAHAM-DIXON, CARAVAGGIO: A LIFE SACRED AND PROFANE 214 (W.W. Norton & Company, 2012)
[311] *Id.*
[312] Id.
[313] *See* ANDREW GRAHAM-DIXON, CARAVAGGIO: A LIFE SACRED AND PROFANE 215 (W.W. Norton & Company, 2012)

## C. Originality Applied - from Traditional Creativity to AI-Generated Works

Applying the *Feist* standard above, it becomes clear that the model users meet the originality requirement when they use them to generate images. They independently create Works that show minimal creativity when they carefully craft detailed prompts to describe their envisioned scenes and iteratively refine the GenAI's output with additional tools until it aligns with their thought. The specificity and depth of these prompts reflect the users' extensive and deliberate conceptualization, showing the inherent originality in their creative process – Jason Allen changes hundreds of prompts and personally picks all elements lightening, composition, styles; the traditional AI artist described above in the first example uses Stable Diffusion for an initial pose, then keeps refining the image until he reaches something satisfactory. The conceptualisation, ideation, and iterations all belong to the artist; the GenAI models are simply a tool for the artists to use to arrive at the ideal picture.

A potential objection to considering the user of a GenAI model as the author of the resulting image might be the unpredictability in the AI's output, as the user cannot anticipate the exact result of a given prompt. However, this perspective is flawed. The presence of unpredictability in the creative process does not strip the creator of their authorship. Many artists intentionally incorporate elements of chance into their work, yet Copyright law still recognizes their authorship.

For example, consider Tim Knowles's *Wind Walk* series. In these works, Knowles attaches drawing tools to trees and lets the wind determine the course of the drawings.[314] While the wind's direction and intensity are outside Knowles's direct control, his creative intent is unmistakably evident. He carefully orchestrates the entire project, from visually capturing the movement of the wind to selecting and positioning the drawing implements.[315] This deliberate setup shows his control and originality; it also shows that, despite the randomness inherent in his method, the resulting work is a direct manifestation of his vision. His method of harnessing nature's unpredictability into a coherent artistic statement is a direct testament of how creative choices can transform random, uncontrollable forces into distinctive art.

Similarly, Jackson Pollock, a pioneer of abstract expressionism, incorporates an element of randomness into his artistic process. Known for his drip painting technique, he masterfully blends chaos with order when he splashes paint on canvas. For example, in *Autumn Rhythm*, he pours, drips, dribbles, scumbles, flicks, and splatters the thinned paint to unprimed, unstretched canvas.[316] This seeming randomness by no means diminishes his control over the resulting image. On the contrary, it shows his ability to harness unorthodox means to create a work that is uniquely his.

When artists incorporate GenAI models into their creative processes, the unpredictability or the model's interpretation of commands doesn't detract from their artistic authority. If the generated output doesn't align with their vision, artists have the option to tweak their prompts and modify the resulting images. They play an active role in the creation process, continually refining the AI's output. Consequently, the final artwork isn't the result of an independent, automatic, or arbitrary process. Rather, it reflects the artist's creativity and their skill in utilizing external forces to craft something uniquely theirs.

---

[314] *See* 21st Century Digital Art, http://www.digiart21.org/art/path-of-least-resistance-and-seven-walks-from-seven-dials (last visited March 3, 2024)

[315] *Id.*

[316] *See* Rosie Lesso, *How Did Jackson Pollock Paint Autumn Rhythm*, The Collector (Jun 16, 2022), https://www.thecollector.com/how-did-jackson-pollock-paint-autumn-rhythm/

## IV. Registration Proposal: What to Include

In the Guidance, the Copyright Office asks the applicants to identify and exclude the parts that are generated by AI. It also requires the applicants to disclose and explain the use of AI-generated content. In this section, I argue that this requirement is unreasonable. As mentioned above, in a coherent picture, whether certain elements are edited depends more on its relation to the other parts of the body than on its own. The fact that some parts of the picture aren't edited by the Office doesn't mean that the GenAI model is the creator. This might simply mean that the human author finds them fitting for the image and in line with their vision, which in turn reflects the originality they bring to the work. Based on this, I suggest that the Copyright Office, instead of requiring applicants to explain the extent of their modifications, could simply ask whether the author's submission includes AI-generated materials and, if so, if AI is integrated in the workflow.

This section is divided into three parts. The first part examines the current Guidance in detail, identifying the themes that the Office is focusing on. The second part criticizes these themes; it argues that the Office's approach to distinguish between human-created and machine-generated content fails to acknowledge the nuanced interaction in AI-assisted creative processes. In this context, an author's decision to retain AI-generated content can be a significant artistic choice, not a relinquishment of creative control. In the third part, I offer an alternative proposal - instead of requiring applicants to identify specific elements created by GenAI models, the Copyright Office could provide a checklist to ask applicants to clearly indicate and confirm the substantial involvement of GenAI in the creative process. Such an approach would not only streamline the process but also mark a pivotal step towards updating current copyright laws to accommodate emerging technologies.

## A. Current Guidance

The Guidance currently 1) requires the applicants to disclose the use of GenAI in their creative process, 2) stipulates that copyright protection would only extend to the parts created by humans, 3) requires that applicants exclude the elements generated by GenAI that are more than de minimis, 4) mandates that applicants shall revise their applications for GenAI if they haven't done so.

Requirement 1: Duty to disclose.

The Guidance requires that the "*applicants have a duty to disclose the inclusion of AI-generated content in a work submitted for registration.*"[317] The applicants shall not hide their integration of generative models in their creative process from the Copyright Office, as "*such disclosures are information regarded by the Register of Copyrights as bearing upon the preparation or identification of the work or the existence, ownership, or duration of the*

---

[317] Copyright Registration Guidance: Works Containing Material Generated by Artificial Intelligence, 88 Fed. Reg. 16,190 (Mar. 16, 2023) (to be codified at 37 C.F.R. § 202)

*copyright.*"[318] While they are doing so, the applicants shall "*provide a brief explanation of the human author's contributions to the work.*"[319] The Office needs to know what is contributed by whom - whether it is the human author or the generative models.

<u>Requirement 2: Rights of Applicants</u>

Copyright protection will only be extended to the parts of the Work created by humans. Applicants "*who use AI technology in creating a work may claim copyright protection for their own contributions to that work.*"[320] The protection extends only to their own, personal, human contribution. It doesn't cover the parts created by the machine. It believes that "*an applicant who incorporates AI- generated text into a larger textual work should claim the portions of the textual work that is human-authored.*"[321]

<u>Requirement 3: Guidance for Applicants who Incorporate AI-Generated Materials in Their Works</u>
The Copyright Office seems to divide the applicants into two groups - those who creatively arrange the elements in the picture, and those who use the generative models to create a portion of picture that is more than de minimis. While humans can creatively select and arrange elements created by humans and machines, the portion generated by machines can only be minimal. The portion that is more than the minimal shall be excluded. In both cases, the applicants shall use the Standard Application, and in it identify the authors and provide a brief statement that describes their contribution that is not created or assisted with a machine.

As the Guidance states, for the applicant who "*creatively arranges the human and non-human content within a work,*" they should claim "*Selection, coordination, and arrangement of [describe human-authored content] created by the author and [describe AI content] generated by artificial intelligence*" in the "*Author Created*" section.[322] For the applicants who incorporate an amount of AI-generated content that is more than de minimis in their work, the part that is created by the model should be explicitly excluded from the application. This "*may be done in the 'Limitation of the Claim' section in the 'Other' field, under the 'Material Excluded' heading.*"[323] In the meanwhile, applicants should also briefly describe the AI-generated content, "*such as by entering '[description of content] generated by artificial intelligence.*'"[324]

<u>Requirement 4: Guidance to Previous and Current Applicants for Application Submission</u>
The Copyright Office advises that "*applicants who have already submitted applications for works containing AI-generated material*" must disclose this information in their application to ensure the effectiveness of their registration.[325] This requirement also applies to applicants whose applications are currently pending before the Office and applicants whose applications have already been processed and resulted in a registration.[326] For those with pending applications, they are advised to contact the Copyright Office's Public Information Office and report that their application omits the fact that the work contains AI-generated material.[327] Meanwhile, applicants with registrations already processed should update the public record by "submitting a supplementary registration."[328]

---

[318] *Id.*
[319] *Id.*
[320] *Id.*
[321] *Id.*
[322] *Id.*
[323] *Id.*
[324] *Id.*
[325] *Id.*
[326] *Id.*

As a principle, applicants should not list an AI technology or the company that provided it as an author or co-author simply because they used it when creating their work. If they are unsure of how to fill out the application, they may simply provide a general statement that a work contains AI-generated materials.

## B. Criticism of Current guidance

I provide two criticisms here: the first addresses the impracticality of excluding significant machine-generated contributions; the second focuses on the theoretical challenge of defining what constitutes a minimal contribution. I argue that the current policy fails to recognize that authors can exercise selective and creative control over extensive AI-generated content. Consequently, a more progressive policy that reflects this understanding is warranted.

Criticism 1: Authorship is more about having control over the creative process than the origin of the content.

The Office's guideline to exclude any AI-generated component exceeding minimal input is unrealistic. Authorial control isn't solely about whether elements are machine-generated. It hinges on the author's ability to review the entire piece and deem these elements congruent with the overall vision. If the author can confirm that even the AI-generated parts are consistent with their conceptualization, and they have personally edited, selected and arranged the parts, they shall meet the requirement for originality.

Criticism 2: it is theoretically impossible to determine what is minimal in machine generated work

The Copyright Office requires in the Guidance that any part generated by AI that is more than the minimum shall be excluded. This is theoretically impossible - in a coherent picture, determining what constitutes a minimal contribution is challenging. For instance, are the style and theme of an image considered minor or major elements? How about the selection and arrangement of all elements generated by AI?

Take Jason Allen's Space Opera Theater series as an example. In these works, GenAI plays a crucial role generating elements such as the Victorian dresses, thematic nuances, lighting, helmets, stylistic touches, and the overall ambiance. By objective standards, AI creates a significant part of the pictures. But in practice, it is Allen that maintains the full control over the conceptualization, arrangement and selection, evaluation, and confirmation. GenAI's role in the creation is intimately intertwined with the human author's engagement. To exclude the portion generated by the models would devalue Allen's contribution.

To summarize, denying copyright registration based solely on the quantity of machine-generated elements overlooks the collaborative essence of such creative works. The existing policy does not fully take into account the sophisticated and interactive approach authors use to integrate AI into their creative process. It neglects the fact that authors can exert selective and innovative control over substantial AI-generated content. By strictly interpreting significant AI contributions as a lack of authorial control, the policy may hinder innovation and undervalue the creative contributions of human authors who work with AI in their artistic endeavors. A more nuanced approach that acknowledges the cooperative relationship between human creativity and AI in art creation is crucial for an equitable and progressive copyright policy.

---

[327] *Id.*
[328] *Id.*

## C. Proposal

Rather than focusing on the quantity of AI-generated content that can be allowed within a Work for registration, the Office should acknowledge that human authors can exert significant control over AI-generated content. The primary concern is not the volume produced by humans, but rather their ability to maintain control over it.

Therefore, the Office might consider issuing a simple checklist in the Application that asks - is AI used in the creative process? If yes, is it fully integrated in the workflow for whom registration is sought? This approach would make the application process more efficient and leverage market forces to discourage applicants who haven't made significant creative investments from applying.

## V. Policy Considerations for Giving Authors Copyright

Several arguments are in favor of the position granting authorship to GenAI model users who incorporate the technology in their creation process. First, recognizing model users as authors aligns with the constitutional objective of advancing useful arts and sciences, especially as GenAI has shown great potential to transform the artistic landscape. Second, this approach moves us away from the reactive proclivity of crafting regulations in response to technological progress to aiming to establish flexible frameworks and guidelines that can evolve with future developments. Third, acknowledging users' significant contributions to public discourse would encourage more intimate engagement and participation in the wider social discourse about technology's role in society.

### A. Not Giving Model Users Authorship Fails to Fulfill the Constitutional Goal of Promoting Useful Arts and Sciences

Art. 1. Sec. 8, Clause 8 of the Constitution authorizes Congress to promote useful arts and sciences by granting authors and inventors exclusive rights to their works and discoveries for a limited period.[329] Since the introduction of text-to-image generators like Midjourney and Stable Diffusion in 2022, and the more recent launch of the text-to-video generator, Sora, artists have recognized GenAI's potential to redefine creativity and serve as an invaluable tool for artistic expression. This technology has been hailed for revolutionizing content creation, streamlining repetitive tasks, and allowing artists to concentrate on the essence of creation. It also enables individuals without formal artistic training to produce works they previously could not, thereby fostering a more inclusive and democratic artistic landscape. As a result, the Office's refusal to register AI art goes against the constitutional goal of promoting the useful arts and sciences.

---

[329] U.S. CONST. art. I, §8

## B. Setting the Guidelines for Future Regulations

In addition to advancing the constitutional goal of promoting useful arts and sciences, another crucial reason for the Copyright Office to recognize individuals who extensively use GenAI in their creative processes as authors is to shift the focus from merely pursuing technological advancements to establishing a flexible framework that can evolve over time.

Technology is evolving at an unprecedented pace. Since 2022, AI has expanded into areas such as medical decision-making,[330] mental health support,[331] financial advising,[332] and real estate,[333] leaving us constantly amazed by its advancements and perpetually trailing behind major tech firms. Given that each of these innovations has the potential to redefine established norms within their respective fields, there is a risk of falling into a reactive pattern of crafting regulations in response to technological advancements. However, law, with its traditional nature, does not evolve as swiftly as modern technology. The legal system should focus on creating a foundational regulatory framework rather than outright rejecting new technologies. This approach acknowledges the importance of these advancements and takes a proactive step by establishing guiding principles instead of merely keeping pace with technological progress.

## C. Giving Model User Authorship Encourages Them to Help Shape Social Discourse

Granting authorship to model users who incorporate GenAI in their workflow creates a more inclusive and democratic digital environment by encouraging them to actively participate in shaping the narratives and discourses that are fed back into the training data, thereby countering the pervasive influence of big technology companies.

Big technology companies are constantly molding our preferences and behaviors. From shaping body image ideals on Instagram to influencing voting patterns on social media platforms like Facebook, we find ourselves in echo chambers that reinforce our pre-existing beliefs and preferences, guided by algorithms that nudge us towards specific actions through notifications, ratings, and rewards.[334] With AI's expansive databases and the vast amount of content it scrapes from the web, this influence would only be amplified. It'd be easy to imagine AI systems, by generating content based on their training data, contributes to the collective consciousness of the society while significantly swaying public opinion.[335]

---

[330] *See generally*, Shrug A. Alowais et al., *Revolutionizing Healthcare: the Role of Artificial Intelligence in Clinical Practice,* BMC Medical Education 23 (2023) https://bmcmededuc.biomedcentral.com/articles/10.1186/s12909-023-04698-z#:~:text=AI%20can%20be%20used%20to,patient%20care%20across%20healthcare%20settings. (AI can be used to diagnose disease, develop personalized treatment plans, and assist clinicians with decision making)

[331] *See generally*, Zilin Ma et al., *Evaluating the Experience of LGBTQ+ People Using Large Language Models Based Chatbots for Mental Health Support* (Feb 14, 2024) (unpublished manuscript), https://arxiv.org/pdf/2402.09260.pdf (suggesting that LGBTQ+ individuals are increasingly turning to chatbots powered by large language model to meet their mental health needs).

[332] *See* Jeff Spiegel, *How Advisors are Increasing Efficiency and Impact with AI,* Blackrock Advisor Center (Aug 30, 2023), https://www.blackrock.com/us/financial-professionals/insights/how-advisors-use-ai

[333] *See* Gautam Raturi, *How Artificial Intelligence is Changing the Real Estate Industry*, Medium (Sep 5, 2023) https://medium.com/coinmonks/how-artificial-intelligence-is-changing-the-real-estate-industry-d1ead18bc1fd#:~:text=For%20real%20estate%20experts%2C%20AI,to%20evaluate%20real%20estate%20data. (For real estate experts, AI market analysis can be helpful to get information on the real estate market and use that insight to make judgments such as predicting market trends, property valuations, and demand for particular types of properties)

[334] *See* Sandra C Matz et al., *Psychological Targeting as an Effective Approach to Digital Mass Persuasion*, PNAS (2017), https://www.ncbi.nlm.nih.gov/pmc/articles/pmid/29133409/ (People are exposed to persuasive communication across many different contexts: Governments, companies, and political parties use persuasive appeals to encourage people to eat healthier, purchase a particular product, or vote for a specific candidate; *see also* Sandra C Matz et al., *Using Big Data as a Window into Consumers' Psychology*, Current Opinion in Behavioral Sciences (2017),

Moreover, it's well-documented that big tech often marginalizes already disadvantaged groups and perpetuates stereotypes by categorizing and sorting people and information in a way that can homogenize the digital landscape and sidelining diverse voices and perspectives.[336] For example, Tumblr's 2018 decision to ban "adult content" has erroneously flagged many transition-related posts; YouTube's classification of LGBTQ+ content as "adult" also alienated these communities.[337] Such biases are infiltrated and reinforced on a daily basis.

As a result, legally recognizing the contributions of model users could act as a significant countermeasure. The concept is that when users input their ideas, beliefs, and experiences into these systems, and the generative models then use this content to create new information, their contributions are recognized in the output. When the legal framework gives such content creators authorship, it effectively returns agency to the individuals. It democratizes content creation by shifting from a top-down model of information dissemination to one that is driven by users. This approach provides a balance to the narratives dominated by big tech companies and changes the power dynamics within the digital landscape.

Therefore, while the transformative capabilities of AI in reshaping public discourse and personal decision-making are undeniable, allowing users of AI models to be recognized as authors would empower them to actively participate in the world of technology, instead of merely being data sources for corporations.

## VI. Conclusion

In conclusion, the Copyright Office does not fully grasp the technology's nature by suggesting that the GenAI models operate independently and automatically. It also misconceives the interaction between users and the models by comparing user instructions to commissions for artists. Instead of requiring applicants to specify which parts of their work are generated by AI, the Office should simplify the process. Applicants should simply indicate whether they have incorporated GeneAI into their workflow. This approach is not only more efficient but also marks a step towards acknowledging technological advancements, making the system more adaptable to contemporary realities. It aligns with the constitutional aim of advancing the arts and sciences and allows the public to mitigate the

---

https://www.sciencedirect.com/science/article/pii/S2352154617300566 (Big Data offer a cheap, extensive source of consumer information for marketers) influence of major tech companies by contributing to the discourse through their input in the training data. It encourages us to revisit the questions - What role does copyright law play in our rapidly evolving world? How should copyright law adapt to unforeseen technological advancements? And how do we understand the relationship between technology and society?

[335] *See* The Tech Cat, *Artificial Intelligence and Its Use in Manipulating Public Opinion*, Medium (Mar 23, 2023) https://thetechcat.medium.com/artificial-intelligence-and-its-use-in-manipulating-public-opinion-5547beea8684

[336] *See* Olga Akselrod, *How Artificial Intelligence Can Deepen Racial and Economic Inequalities*, ACLU (July 13, 2021) https://www.aclu.org/news/privacy-technology/how-artificial-intelligence-can-deepen-racial-and-economic-inequities (ample evidence of the discriminatory harm that AI tools can cause to already marginalized groups)

[337] *See* Zilin Ma et al., *Evaluating the Experience of LGBTQ+ People Using Large Language Model Based Chatbots for Mental Health Support* (Feb 14, 2024) (unpublished manuscript), https://arxiv.org/pdf/2402.09260.pdf#:~:text=LGBTQ%2B%20participants%2C%20who%20often%20faced,negative%20backlash%20or%20being%20outed.